# Chemical speciation and source apportionment of ambient $PM_{2.5}$ in New Delhi before, during, and after the Diwali fireworks

Chirag Manchanda [a, †], Mayank Kumar [a,*], Vikram Singh [b, *], Naba Hazarika [c], Mohd Faisal [b], Vipul Lalchandani [d], Ashutosh Shukla [d], Jay Dave [e], Neeraj Rastogi [e], Sachchida Nand Tripathi [d,*]

a. Department of Mechanical Engineering, Indian Institute of Technology Delhi, New Delhi, India

b. Department of Chemical Engineering, Indian Institute of Technology Delhi, New Delhi, India

c. Department of Applied Mechanics, Indian Institute of Technology Delhi, New Delhi, India

d. Department of Civil Engineering, Indian Institute of Technology Kanpur, Uttar Pradesh, India

e. Geosciences Division, Physical Research Laboratory, Ahmedabad, India

## Abstract

Diwali is among the most important Indian festivals, and elaborate firework displays mark the evening's festivities. This study assesses the impact of Diwali on the concentration, composition, and sources of ambient $PM_{2.5}$. We observed the total $PM_{2.5}$ concentrations to rise to 16 times the pre-firework levels, while each of the elemental, organic, and black carbon fractions of ambient $PM_{2.5}$ increased by a factor of 46.1, 3.7, and 5.6, respectively. The concentration of species like K, Al, Sr, Ba, S, and Bi displayed distinct peaks during the firework event and were identified as tracers. The average concentrations of potential carcinogens, like As, exceeded US EPA screening levels for industrial air by a factor of ~9.6, while peak levels reached up to 16.1 times the screening levels. The source apportionment study, undertaken using positive matrix factorization, revealed the fireworks to account for 95% of the total elemental $PM_{2.5}$ during Diwali. The resolved primary organic emissions, too, were enhanced by a factor of ~8 during Diwali. Delhi has encountered serious haze events following Diwali in recent years; this study highlights that biomass burning emissions rather than the fireworks drive the poor air quality in the days following Diwali.

**Keywords:** Diwali; Fireworks; Source Apportionment; $PM_{2.5}$; Delhi

## 1. Introduction

The enhancement of heavy metal concentrations in the environment poses a severe threat to the ecosystem and human health, from impacting the microbiological balance in soils (Barbieri, 2016) to an adverse effect on human health (Singh et al., 2011). Intensified metal loading in aerosols, especially fine particulate matter ($PM_{2.5}$), has been a pressing concern considering the increased risk of exposure and ease of bioaccumulation (Tchounwou et al., 2012). Past studies have noted that exposure to high metal concentrations is detrimental to human health, such as increased neurotoxic impairment, respiratory ailment, and carcinogenic impact from metals like Mn, Cu, Pb, Cr (Santos-Burgoa et al., 2001; Wang et al., 2006).

[*] Corresponding Authors
E-mail addresses : kmayank@mech.iitd.ac.in (M. Kumar); vs225@chemical@iitd.ac.in (V. Singh); snt@iitk.ac.in (S.N. Tripathi)
[†] Now at: Department of Civil and Environmental Engineering, University of California, Berkeley, CA, USA

Advances in measurement technology and increased use of statistical tools to apportion the particulate-bound metallic species to real-world sources like traffic emissions, industrial plumes, and mining operations have enabled policymakers to make more informed decisions (Csavina et al., 2011). However, relatively less attention has been drawn towards episodic anthropogenic activities, leading to much more metalliferous particulate matter concentrations in ambient air. Fireworks are often an essential component in recreational activities, celebrations, and festivities across the globe, the New Year's Eve, Las Fallas festival in Spain, the Lantern Festival and Chinese New Year in China, and Diwali in India (Moreno et al., 2007).

Along with particulate matter rich in metallic elements, fireworks are also known to cause high levels of black carbon and generate gaseous pollutants like carbon monoxide, along with oxides of sulfur and nitrogen (Barman et al., 2008). Diwali is an Indian cultural festival celebrated every year around October-November in most parts of India. The celebrations are prominently marked by fireworks displays and firecracker burning during the night. Multiple studies in the past have worked to delineate the after-effects of these festivities on ambient air quality. Babu and Moorthy, (2001) studied the effects of fireworks during Diwali in Thiruvananthapuram, India, and recorded a three-fold increase in the mass concentration of atmospheric black carbon during Diwali when compared with ambient levels during regular days. Goel et al., (2021) observed a similar magnitude of enhancement in ambient BC levels during Diwali 2020 in Delhi. However, the study found this rise to be transient, with the peak concentration observed around midnight plummeting to around one-third by the following day.

Kulshrestha et al., (2004) were the first in India to study variation in particle-bound metal concentration engendered by Diwali and found Ba, K, Al, and Sr concentrations to increase by 1091, 25, 18, and 15 times the typical daily values in Hyderabad. Barman et al., (2008) based their study in Lucknow, India, investigating the effect of Diwali on ambient air quality and reported Cu and Ni to rise by 272% and 149% following the fireworks event. Chatterjee et al., (2013) reported K, Mg, and Ca to increase by a factor of 30, 20, and 14 times in terms of 12-hr averaged concentration levels during Diwali night. Rastogi et al., (2019) documented that the organic aerosols were the dominant contributor (~85%), followed by sulfate (~13%), chloride (~4%), nitrate (~4%), and ammonium (~2%) to non-refractory $PM_1$ on the Diwali night in Ahmedabad, India. Kulshreshtha et al., (2021) investigated the trace metal concentrations in ambient air during the Diwali period in Prayagraj, India, over 2 years (2018-19), and noted substantial enhancements in levels of Pb, Cu, Mn, Ni, and Cr during Diwali for both the years. The abundance of Pb, Mn, and Ni (in addition to Fe and Zn) in elemental PM during Diwali was also noted by Khobragade and Ahirwar, (2022) for their study focusing on the impact of Diwali fireworks on ambient air quality in Raipur during 2018.

Delhi is regarded as one of the world's most polluted cities (Cropper et al., 1997; IQAir World Air Quality Report, 2018), and despite the increasing intensity of pollution levels following Diwali in recent years (Schultz and Raj, 2019), the number of studies focusing on the impact of Diwali in Delhi is found to be very limited. Sarkar et al., (2010) studied the effect of Diwali on the ambient air quality in Delhi and found Ba, K, Sr, and Na in $PM_{10}$ to increase by a factor of 264, 18, 15, and 5, respectively, based on 24-hr averages. Perrino et al., (2011) attributed the rise in ambient $PM_{2.5}$ levels in Delhi to a commensurate increase in Mg, Al, Ti, V, Cr, Mn, and Ga concentrations during Diwali.

Kulshrestha et al., (2014) made a first attempt at characterizing the elements measured during the Diwali event into principal factors using PCA to segregate and quantify firecrackers' contribution as a separate factor. During the regular days with no firework displays, the main components were identified as crustal, vehicular, and industrial, while during the fireworks event, vehicular and firecrackers emerged as the predominant sources. Kotnala et al., (2021) set up measurement sites at 7 locations across Delhi, to study the spatial variability in firework emissions along with the impact of fireworks on the chemical composition of $PM_{2.5}$ at each site during Diwali. The study employed daily filter measurements (one sample per day) at all sites during $13^{th}$ – $30^{th}$ October 2017, these samples were further processed to resolve the organic carbon (OC), elemental carbon (EC), water-soluble inorganic ions (WSIIs) and levoglucosan fractions in the sampled ambient PM. While each PM constituent displayed high variability across different sampling locations, the largest increment was observed for $K^+$ (15.58 fold) and $SO_4^{2-}$(5.1 fold), along with ~30% enhancement in OC on the Diwali Day.

In a more recent work, Yadav et al., (2022) studied the size-resolved elemental distribution in ambient aerosols during the Diwali firework displays in Delhi in 2018. They observed that the enhancement in elemental contribution to total particulate matter was significantly higher in the quasi-ultrafine particulate (q-UFP) fraction (0.25 -0.5 μm) compared to the coarser particles. During the firework emissions, the elemental concentrations in the q-UFP size bin were dominated by As (32.7%) followed by Pb (23.46%), Cu (23.45%), Al (22.9%), Mg (21.98%), V (20.81%), Cd (19.73%), K (18.11%), and Mn (14.07%).

However, most of the studies noted above are limited by the time resolution of their analysis, mainly due to the offline methods employed for processing the samples. Aerosol concentrations averaged over larger intervals often undermine the peak pollutant levels, which become of greater importance while analyzing episodic events like Diwali. In addition to quantifying peak concentrations, higher time resolutions are also essential to better analyze the health risks to the population exposed to firework emissions based on the time and pollutant concentrations associated with the exposure. The present study overcomes these limitations by employing an online XRF-based near real-time metal monitor for highly resolved continuous measurement and characterization of ambient $PM_{2.5}$. The study is placed towards the end of October, enabling us to analyze the impact of Diwali and compare it with the ordinary days, with no firework events. The present study is designed to meet the following research objectives:

1. The study employs measurements of total $PM_{2.5}$, the inorganic ions (sulfate, nitrate, and ammonium), and black carbon concentrations, along with the elemental and organic fraction of total $PM_{2.5}$, to assess the impact of the fireworks on total $PM_{2.5}$ and its composition.
2. The study focuses on the source apportionment of the elemental and organic $PM_{2.5}$ data to assess the time variation of sources contributing to organic and elemental $PM_{2.5}$ around the Diwali event and hence, quantify the impact of Diwali on ambient air quality in Delhi.
3. The study employs not one but two Xact 625i monitors located at different sites, around 15 km apart, in Delhi. The source apportionment analysis is applied independently to the data from both these instruments, in order to evaluate the sources' local variability contributing to elemental $PM_{2.5}$ during the firework event and the regular days.

## 2. Materials and Methods

## 2.1 Sampling Location and Study Design

The present study is based in the National Capital Territory (NCT) Delhi and spans over 12 days from 20th October 2019 to 31st October 2019, with the Diwali festival on 27th October 2019. To account for the immediate effect of the firecracker burning, we focus on a six-day interval around the Diwali fireworks event and subdivide it into three phases Pre-Diwali (PD) (25th October 2019 24:00 IST – 27th October 2019 18:00 IST), During Diwali (DD) (27th October 2019 18:00 IST – 28th October 2019 6:00 IST), and Following Diwali (FD) (28th October 2019 6:00 IST – 31st October 2019 24:00 IST).

The DD phase was structured to approximately coincide with the start of the firecracker burning on the evening of the Diwali day, leading to a surge in $PM_{2.5}$ concentrations during the night, and end around the point when the pollution levels fall back to nearly normal levels as the boundary layer height starts to increase in the morning following the Diwali night. However, in the case of the elemental $PM_{2.5}$ (denoted using $PM_{2.5}^{el}$) levels and its apportioned sources discussed in sections 3.1 and 3.2, respectively, the enhancement in concentration in the DD phase is referenced against an extended Pre-Diwali phase (ePD) (20th October 2019 24:00 IST – 27th October 2019 18:00 IST).

As discussed in section 1, sampling was conducted at two sites approximately 15 km apart. Figure 1 presents the spatial location of the two measurement sites employed in the present study.

a) The first site was the Indian Institute of Tropical Meteorology campus, Delhi (IITMD), located in Central Delhi (28.6304° N, 77.1751° E). The location is surrounded by an expansive green stretch of the Pusa Forest, and no primary industrial sources are located within a 5 km radius. The instruments were placed in a temperature-controlled facility on the 2nd floor of the complex—the sampling inlet of the instruments extended to the rooftop around 15 m from ground level.
b) The second site was at the Indian Institute of Technology, Delhi (IITD) campus, located in South Delhi (28.5450° N, 77.1926° E). The instruments were housed in a temperature-controlled laboratory on the top floor of a four-story building on campus. The nearest local emissions source is an arterial road outside the campus, located about 150 m from the building.

## 2.2 Instrumentation

The present study employs the Xact 625i Ambient Metal Monitor, installed with a $PM_{2.5}$ inlet for sampling, at both sites to measure $PM_{2.5}^{el}$ concentrations, with a half-hourly time resolution for the entire study period.

From the PD to the FD phase, the Xact 625i at the IITMD site was supplemented with a multi-channel Aethalometer (Magee Scientific Model AE33, Berkeley, CA) for measuring black carbon at a one-minute time resolution, along with a High-Resolution Time-of-Flight Aerosol Mass Spectrometer (HR ToF AMS, Aerodyne Inc., MA, USA), measuring the non-refractory fraction of $PM_{2.5}$ (NR-$PM_{2.5}$). The HR ToF AMS measures the organic fraction of total $PM_{2.5}$ along with inorganic ions like sulfate, nitrate, ammonium (SNA), and chloride. A time resolution of two minutes was used for the AMS.

While both the AMS and Xact measure Cl concentrations, the chlorine fraction reported in this study is based on the Xact 625i values, as AMS only measures the non-refractory ions and consequently the total Cl reported by AMS is consistently found to be less than or equal to the Xact reported levels ( Manchanda et al., 2021).

Further details on the instrumentation and data quality assurance or quality control (QA/QC) checks for each instrument are provided in the supplementary information (SI) section S1.

## 2.3 Source Apportionment using PMF

Positive Matrix Factorization (PMF) is a multivariate factor analysis technique widely used for source apportionment of particulate matter constituents (Paatero and Tapper, 1994a; Ulbrich et al., 2009; Vossler et al., 2016). The PMF model attempts to best express the covariance of the multivariate input data set in terms of a set of time-invariant factor/source profiles and their time-varying contribution, as shown in eq. (1):

$$\mathbf{X} = \mathbf{GF} + \mathbf{E} \quad \ldots\ldots. (1)$$

Where **X** is an $m$ x $n$ input data matrix with a measured concentration of $n$ species at $m$ time steps, **G** is a $m$ x $p$ fractional contribution matrix quantifying the fractional contribution of $p$ source profiles at each of the $m$ time steps. **F** is a time-invariant matrix representing a set of $p$ factor profiles/vectors constituted by $n$ elements, where the user selects the number of factors, i.e., p. **E** is a $m \times n$ matrix of residuals corresponding to the difference ($e_{ij}$) between each element of the input matrix **X** and the product of matrices **G** and **F**.

The PMF algorithm was implemented using EPA PMF 5.0, which is, in turn, built on the ME-2 solution model (Paatero, 1999). Further details on the model are discussed in past studies (Paatero, 1997; Paatero and Tapper, 1994b). Further details w.r.t the present study regarding PMF input preparation, factor selection, and uncertainty quantification have been reported in SI section S2.

# 3.Results and Discussions

Section 3.1 presents the variation in total $PM_{2.5}$ in the Diwali period and the change in its composition during each phase from PD to FD. We further pivot our discussion to the constituents of $PM_{2.5}^{el}$ augmented during the DD phase and the possible role of fireworks in enhancing these species, which are further observed to mark the firework event. Section 3.1.1 discusses the potential health risks associated with firework-related emissions during Diwali. To quantify the contribution of firework displays to the organic and elemental fractions of $PM_{2.5}$, we perform a source apportionment study using PMF presented in section 3.2.

## 3.1 Variation in $PM_{2.5}$ and its constituents during Diwali

As discussed in section 2.1 earlier, from the PD up to the FD phase, we commissioned an extra set of instruments at the IITMD site to get a broader picture of the firecracker event's impact on total $PM_{2.5}$. We must note that each instrument is designed to measure partial rather than total $PM_{2.5}$, measuring each of the elemental, organic, inorganic ions, and black carbon components. While none of these instruments can measure refractory ions like sodium, magnesium, oxides, and fluorides, the residual mass we obtained comparing the instrument total $PM_{2.5}$ and total $PM_{2.5}$ from a Beta Attenuation Monitor (BAM) is negligible. The instrument total was regressed against the BAM measurements, and an appreciable correlation (Pearson R = 0.953) between their time variation was recorded (figure S1).

Figure 2a) presents the time variation of the instrument total $PM_{2.5}$ from the PD to the FD phase; we note that the firework event results in a sharp rise in total $PM_{2.5}$, with the average concentration in the DD phase (around 192 $\mu g/m^3$) increasing by 408% compared to the PD phase. The peak concentrations observed within the DD phase (~ 606 $\mu g/m^3$) are more than 16 times (or 1500 % rise) w.r.t the PD phase. After the fireworks event, the FD phase shows a significant dip in the particulate concentrations; however, the average particulate concentrations remain higher than the PD phase by a factor of ~2.

Figure 2(b) shows the variation in the composition of the $PM_{2.5}$ across the 3 phases. We observe that similar to the total $PM_{2.5}$, each constituent fraction increases during the DD phase; however, each component's percentage rise is highly disproportionate. The organics, BC, and SNA fractions increase by approximately 1.6, 2.3, and 2.5 times the PD concentrations. The chloride fraction (measured using Xact 625i) increases by 640% or 7.4 times during the DD phase, while the metallic fraction rises to around 19.4 times the PD value. Most fractions reduce considerably during the FD phase but remain higher than the PD levels. BC during FD remains about 30% higher than PD, while metals and chloride stay at around 3.5 and 3 times their PD levels. It is essential to note that SNA remains approximately 2.1 times the PD concentrations, reducing by only 15% compared to the average in the DD phase, signaling towards either the lasting impact of the Diwali event on the inorganic ionic concentrations or some other anthropogenic or meteorological cause leading to the increase, this will be discussed further in section 3.2. Again, for the organic fraction (figure 2(b)), it is interesting to note that the levels in the FD phase average around 2.2 times the PD phase and are, in turn, 37% higher than even the DD phase. The behavior of the organic fraction is discussed further in section 3.3 and SI section S3.5 when the organic fraction is subjected to source apportionment to identify possible reasons that lead to this increase.

Figures 2(c-d) present the variation of $PM_{2.5}^{el}$ ($PM_{2.5}^{el}$ = Metals + Cl + S), measured using Xact 625i at both the sampling sites, i.e., IITMD and IITD, throughout the study period, while figures 2 (e - j) present the variation of each element constituting $PM_{2.5}^{el}$ at IITD and IITMD through each of ePD, DD, and FD. At IITMD, $PM_{2.5}^{el}$ rises by 12.4 times compared to ePD and stays at around 2.8 times the ePD value in FD, while at IITD, $PM_{2.5}^{el}$ increases to 12 times the ePD levels and remain at about twice the ePD concentrations in FD. However, the average $PM_{2.5}^{el}$ concentration remains ~155 $\mu g/m^3$ during the DD phase at both the IITD and IITMD sites.

In terms of the elemental variation at both the sites during each phase (figure 2 (e-j)), we note that elements like Al, Ba, Cl, K, Sr, and S are recorded in considerable concentrations at both locations during the study period and rise significantly in the DD phase. Aluminum rises by 69.6 and 97.3 times in the DD phase w.r.t ePD at IITMD and IITD sites, respectively, with an average concentration of 44.2 $\mu g/m^3$ (IITMD) and 48.8 $\mu g/m^3$ (IITD) during the DD phase. Similarly, during the DD phase, Ba concentration averages at 15.4 $\mu g/m^3$ (IITMD) and 14.3 $\mu g/m^3$ (IITD), Cl at 11.4 $\mu g/m^3$ (IITMD) and 9 $\mu g/m^3$ (IITD), K at 48.2 $\mu g/m^3$ (IITMD) and 53.2 $\mu g/m^3$ (IITD), Sr at 1.5 $\mu g/m^3$ (IITMD) and 1.1 $\mu g/m^3$ (IITD), and S at 25.3 $\mu g/m^3$ (IITMD) and 23 $\mu g/m^3$ (IITD). In terms of the increment in concentration during the DD phase, Ba rises by 190.4, 123.4 times, Cl by 4.7, 6.5 times, K by 22.5, 19.8 times, Sr rises by 145, 87.2 times, and S by 6.3, 5.1 times at IITMD and IITD sites respectively when comparing the DD to the ePD average concentrations. There are certain species where the concentration was close to the MDL in ePD and rose substantially in the DD phase. At IITMD, Vanadium and Bismuth averaged around 0.7 $ng/m^3$ and 0.4 $ng/m^3$ in the ePD phase while rising to 0.43 $\mu g/m^3$ and 0.13 $\mu g/m^3$ respectively in DD, thus increasing ~608 and ~321 times their ePD concentrations. We notice that while the firecracker burning leads to a considerable rise in the concentration of certain species, other species remain primarily unaffected. Various studies in the past have recognized this disproportionate behavior pointing to the emissions specific to firework events and identified these elements as a markers for the fireworks (Shon et al., 2015; Tanda et al., 2019).

In the present study K, Al, Sr, Ba, S, and Bi were found to display distinct peaks during the DD phase at both our sampling locations (Figure S3 (a-f)) and were recognized as elemental tracers for the firework event. Apart from elemental markers, we also observe a marked trend in BC concentrations (figure S3 (h)) in the DD phase, with the average levels rising to about 2.2 times the PD concentrations, while peak concentrations reach ~5.6 times the average PD levels. We can also take note of the elevated $SO_4$ to $NO_3$ ratios (figure S3 (g)) acting as a marker to the firework event; we observe this ratio to increase by around 2.1 times the PD levels. Further discussions on these tracers and their potential relation to firecracker constituents have been presented in supplementary information (SI) section S3.1.

*3.1.1 Potential health hazards associated with the Diwali fireworks*

It is important to note that, while past studies affirm the toxic nature of firework emissions (SI section S3.2), the carcinogenic/non-carcinogenic health risks associated with the emitted species are primarily based on prolonged exposures at ambient pollutant levels (Lin, 2016). However, in the case of firework events like Diwali, the pollutant levels are much higher, but the duration of exposure is much shorter. Similar concerns have been put forth by several recent review articles focusing on firework-related emissions across the globe (Lin, 2016; Singh et al., 2019).

Some studies have employed traditional risk assessment techniques to quantify the hazardous impact associated with firework-related emissions during Diwali (Sarkar et al., 2010). However, it is important to note that most existing health impact and exposure assessment models are established for long-term average exposure levels, and deal with time scales relevant to average daily exposures/intake. While it is possible to evaluate excess risk based on comparing exposure metrics by selectively including the enhanced averaged levels on Diwali day (Sarkar et al., 2010), such metrics would highly undermine the actual risks associated with short-lived but high-intensity exposures associated with firework events (Lin, 2016; Singh et al., 2019). This is mainly because the epidemiological impact of short-lived high-intensity exposures is still not well established compared to the long-term exposures in the case of health impact assessment (Singh et al., 2019).

Licudine et al., (2012), utilized the EPA Risk-Based Concentration (RBC) tables (or the Generic Regional Screening Levels (RSLs)) to reference against the trace metal concentrations observed during the New Years' Eve fireworks in Pearl City, Hawaii, USA. Table 1 provides a comparison between the elemental concentrations recorded during the DD phase and the EPA generic RSLs (United States Environmental Protection Agency, 2020) for both residential and industrial air, for each of the elements measured for which the RSLs were available. We note that, for metals like Al and Ba, which were recognized as markers for the Diwali event in this study, the 12-hr average concentrations exceeded not only for the residential air quality standards but also the industrial air RSLs. The average levels of Al during the DD phase exceeded the industrial air SL by a factor of ~2 at both the IITMD and IITD sites, whereas the maximum levels reached up to 5.2 and 8.1 times the industrial SL at the IITD and the IITMD sites, respectively.

Similarly, for Ba, the average and maximum levels exceeded the industrial SLs by a factor of 6.5,16 at the IITD site and 7, 29 at the IITMD site. In the case of potential carcinogens like As (Table 1), the 12-hr average concentration at IITMD exceeds the residential RSL by a factor of ~25. In other words, an individual in Delhi inhaled more As in just 12 hours than what the US EPA would have deemed safe over 25 days for residential air. Again, in case of As both the 12-hr average and the maximum levels during the DD phase at both the sampling locations exceeded the industrial air SLs by a factor of 4.8 and 16.1 respectively (taking a mean across the two sites), further signifying the intensity of emissions associated with Diwali fireworks. Similar behavior was displayed by other elemental species like chlorine and lead, with the maximum concentrations exceeding the SLs by more than an order of magnitude. However, in the case of manganese, the observed concentrations exceeded the residential air SL but were under the industrial air SL, while in the case of nickel, selenium, and vanadium, the measured concentrations remain below the residential air SL.

These observations warrant the need for further studies focused on evaluating the health impact of high-intensity, short-duration events like firework displays on human health and quantifying the risks associated with high intensity, limited duration exposure characteristics of such events. Such a work would be of prime importance in the case of Diwali, given that the emissions associated with Diwali, as per the present study, exceed those from similar firework events across the globe by order of magnitude (SI section 3.3).

### 3.2 Source apportionment of elemental $PM_{2.5}$

In order to shed light on the variation in sources contributing to the elemental fraction of ambient $PM_{2.5}$ before, during, and after Diwali, the Xact 625i-based measurements at both the IITMD and IITD sites were individually subjected to source apportionment using PMF. The input dataset was found to be best represented by a seven-factor solution at both the sampling locations, i.e., industrial emissions, coal combustion, vehicular emissions, fireworks, dust-related, biomass burning, and secondary chloride. Further details on factor selection and uncertainty quantification have been provided in SI section S2.3.

The PMF resolved factors were associated with real-world sources based on the species dominating each factor profile (figure S4). A detailed description of the composition and naming of each factor profile at both the measurement sites has been provided in SI section S3.4. This section focuses on the temporal variation and relative contribution of each source to the elemental $PM_{2.5}$ loading at both the measurement sites.

*Industrial Emissions:* The source is marked by high Zn, K, and S levels at both the IITMD and IITD sites (Figure S4). While the industrial emissions factor profiles share similarities at both the measurement sites, some of the contributing species like Cu, As, Br have been found to be local to individual sites; such behavior potentially stems from disparate localized emissions specific to each location and is discussed further in section S3.4.1.

There seems to be no consistent behavior between IITMD (figure 3(a, c)) and IITD (figure 3(b, d)), in terms of the variation from the ePD to FD phase. The factor concentration decreases in the DD phase at IITMD, while it increases at the IITD site. However, the factor concentration in the FD phase is lower than the ePD phase at both sites, which could stem from limited industrial activities during a period of extended festivities (that lasts up to 2-3 days following Diwali). This difference in the time variation of this factor at both sites further supports the hypothesis of localized sources/meteorological conditions affecting the factor profiles. The diurnal behavior (figure S7 (a & b)) of this source at both sites is found to be coherent with each other, and 2 sharp peaks around midnight and early morning (6:00 IST) are observed. The increasing concentrations around midnight may result from the lowering boundary layer height (BLH) at night, while the early morning peak followed by a rapid decline in levels indicates the role of gas-particle partitioning.

*Coal Combustion:* Enriched S and Pb levels mark the coal combustion factor profile at both the measurement sites (figure S4). The time variation associated with coal combustion (figure 3(a, b)) displays peaks of different intensities at both the sites; however, the diurnal variation (figure S7 (c & d)) shows the factor to reach peak values around 6:00 IST at both locations. Similar to the industrial emissions, the diurnality seems to be influenced by the lower temperatures and decreased volatility at night, leading to condensation of aerosols in the particulate phase and rising in concentration until early morning when temperature and, in turn, volatility starts to increase, and the aerosols evaporate back from the particle phase.

*Vehicular Emissions:* Elemental markers such as Mn, Cr, and Ni mark the vehicular emissions source profile at both the measurement sites in the case of the present study (figure S4). Looking at the variation in concentration and contribution of this factor at both the locations (figure 3), we note that the vehicular emissions at both sites decrease by more than 95% on average during the DD phase as compared to the ePD phase and remain lower than the ePD values even in the FD phase. This decrease can be attributed to reduced mobility due to holidays in educational and most commercial establishments from the DD phase to the FD phase. In terms of diurnal behavior, the source resolved at the IITMD site (figure S7(e)) displays distinct rush-hour peaks around 6:00 IST to 9:00 IST in the morning and 18:00 IST to 21:00 IST during the evening. However, at the IITD site (figure S7(f)), the most prominent peak is observed between 3:00 IST to 6:00 IST, indicating the role of heavy motor vehicle traffic possibly due to the vicinity of the site to National Highway (NH) - 48 (around 8 km NW). Other than this, the diurnal behavior is not as distinct as observed for IITMD, while peaks are observed at potential rush hours, i.e., 9:00IST, 16:00 IST, and 19:00 IST, but the characteristic rush hour trend is suppressed by the presence of multiple busy roads around the IITD site and limited (12-day) sampling period.

*Fireworks:* The fireworks factor profile (figure S4) is populated by the tracers linked to firework constituents, as discussed earlier in section 3.1 and SI section 3.1 and 3.4.4. The temporal evolution of the firework emissions at both the measurements sites (figure 3) highlights that firework associated emissions are not only limited to the Diwali day but also appear on a day prior and a day later to Diwali (figure 3(a, b)), this observation corroborates with the fact that in India, it is common to observe limited and scattered firecracker burning events before and after the Diwali day.

It is noteworthy here that the peaks associated with the fireworks source (figure 3(a, b)) may seem inconsistent across the two sites, especially in the FD phase. The authors believe that these disparities in the firework source between the two sites are due to localized effects close to the receptor. In order to ascertain whether the peaks associated with the fireworks source in the ePD and FD phases are actually due to scattered firecracker burning events before and after the Diwali day, or just statistical remnants from the PMF analysis, we attempted to constraint the firework emissions to only the DD phase. The imposed constraints were found to impact the source profiles and variation of other sources and resulted in a rotationally unstable solution; the same is discussed in further detail in SI section S2.3. This indicates that the fireworks peaks in the ePD and FD phases do correspond to actual emissions, although the emission and the intensity are localized to the receptor site.

The average concentration at the IITMD site in the DD phase is around 390 times the ePD values, while the peak concentrations are ~1576 times the average ePD levels, whereas, at IITD, the average concentrations in the DD phase were around 178 times, while the peak concentrations around 332 times the average ePD values. It is interesting to note the change in $PM_{2.5}^{el}$ composition during DD while the average contribution of fireworks to $PM_{2.5}^{el}$ was just 3% at IITMD and 6% at IITD in the ePD phase, it rose to 94% and 92% in the DD phase at IITMD and IITD respectively. While the peak concentrations plummet within ~12 hours following the firework event, the firecracker burning on the day following Diwali and to some extent, the residual firework-related particulate content from the Diwali day cause this factor to account for around 54.3 % of total $PM_{2.5}^{el}$ at the IITMD site and 29.5% of total $PM_{2.5}^{el}$ at the IITD site during the FD phase. However, it's important to note that some technical difficulties disrupted sampling on 29$^{th}$ October at the IITMD site. Thus, the limited measurements in the FD phase at IITMD are partially responsible for the inflation in percentage contribution of the fireworks source at IITMD compared to IITD in the FD phase.

*Dust-Related:* While the dust-related factor is marked by crustal elements Si, Ca, and Fe at both the measurement sites, as discussed in section S3.4.5, the time variation associated with the dust-related factor seems to follow no discernable trend (figure 5(a, b)) or the diurnal profiles (figure S7 (i & j)), at the IITMD or IITD sites. It is evident that this factor remains mostly unaffected by the Diwali event.

*Biomass Burning:* The biomass burning factor profiles associated with both the measurement locations display apparent domination of K and S levels (SI section 3.4.6, figure S4). As discussed earlier in section 3.1, the October to November period in India coincides with the crop harvesting season in northern India and witnesses a rise in biomass/stubble burning events (Jethva et al., 2019). The time variation corresponding to this factor displays a positive trend throughout the study period (figure 3(a, b)). In terms of the diurnality (figure S7 (k & l)), we observe elevated concentrations at both the sites after 18:00 IST until early morning; this behavior may stem from increased emissions from stubble burning events at night, and increased use of domestic biomass for heating purposes due to the cold weather, along with decreasing BLH after sunset. From the variation in percentage contribution (figure 3(c, d)), we note that the factor falls in concentration from the ePD to the DD phase; however, it rises back in the FD phase to a concentration higher than the ePD phase. Past studies (Wang et al., 2012) have proposed using Delta-C (the difference between the Aethalometer measurements at 880nm and 370 nm, respectively) as a marker for biomass and wood burning. In the present study, we note a Pearson R correlation of 0.67 (figure S6 (a)) between the temporal evolution of $PM_{2.5}^{el}$ apportioned biomass burning source and Delta-C.

*Secondary Chloride:* The secondary chloride factor profiles showcase apparent domination of Cl, followed by notable levels of Br (SI section 3.4.7, figure S4). A similar factor has been resolved by some past source apportionment studies based in Delhi. Jaiprakash et al., (2017) observed a similar source profile while studying sources of $PM_1$ levels in Delhi and speculated this factor to emerge as HCl fumes transported from metal processing plants to the northwest of Delhi, reacting with high $NH_3$ concentrations in the region to condense on the particulate phase as $NH_4Cl$ (Harrison et al., 1990; Warner et al., 2017). Recent studies based in Delhi, such as Manchanda et al., (2021) and Gani et al., (2019), reached similar conclusions for the high particulate-phase chloride concentrations observed by them.

In the present study, we note that the secondary chloride factor is enhanced in terms of concentration in the DD phase (figure 3), suggesting some influence of the fireworks event. These enhanced concentrations could result from a scenario where excess aqueous ammonia (Warner et al., 2017) replaces the metallic cation from the primary aerosols released from the combustion of perchlorate oxidants present in the fireworks, as discussed in SI section 3.1. It is quite possible that while some portion of the total chlorine emissions from fireworks, which doesn't react with the excess ammonia, is apportioned to the fireworks-related source, another part that undergoes secondary reactions with ammonia leads to the enhancement of secondary chloride. A similar situation is possible in the case of biomass burning; while KCl is regarded as a primary emission from biomass burning (Li et al., 2003), the biomass burning factor accounts for less than 15% of total chlorine concentrations (figure S4). The increasing biomass burning levels during the FD phase (figure 3 (a, b)) may thus be the reason behind the elevated levels of secondary chloride in the FD phase as compared to the ePD phase at both the sampling sites (2.87 times at IITMD and 4.58 times at IITD). The diurnal variation associated with the factor (figure S4 (m & n)) displays a sharp peak around 6:00 IST. The variation is very similar to the expected diurnal trend for $NH_4Cl$ (Warner et al., 2017). Early morning peaks for $NH_4Cl$ stem from low saturation vapor pressure at lower temperatures at night, leading to condensation of $NH_4Cl$ in the particulate phase, followed by the consequent increase in saturation vapor pressure in the morning responsible for the sharp decline in concentration.

### 3.3 Impact of firecracker burning on ambient $PM_{2.5}$ levels

It is interesting to note that $PM_{2.5}$ or fine mode particulate matter is often associated with longer residence time in the atmosphere, but the total $PM_{2.5}$ levels in the present study (figure 2(a)) display a rapid downfall in concentration following the maxima around midnight in the DD phase. The elemental $PM_{2.5}$ apportioned to fireworks, as discussed in section 3.2, also displays an intense peak around midnight, followed by a steep drop up till 6 am on the following day. In addition to the source apportionment of elemental $PM_{2.5}$, the organic fraction of $PM_{2.5}$ was also apportioned to primary organic aerosol (POA) and oxygenated organic aerosol (OOA) sources, discussed in detail in SI section S3.5. Similar to the total $PM_{2.5}$ and the firework-associated elemental $PM_{2.5}$, the POA source displays a notable enhancement (~7.8 times) in concentration in the DD phase compared to the PD phase (figure S5(b)). On the other hand, while the OOA sources display around a 30% increment during the DD phase w.r.t. the PD phase, it displays a much more pronounced rise (as high as 186% w.r.t the PD levels for OOA-2) during the FD phase. The POA and the BC levels too display a concurrent surge in the FD phase (figure S6(b)) with a Pearson R correlation of 0.89. Again the rise of the Delta-C tracer and biomass burning fraction of elemental $PM_{2.5}$ (figure S6(a)) in the FD phase, as discussed in section 3.2, indicates that the enhancement in total $PM_{2.5}$, as well as the elemental and organic fractions in the FD phase, is primarily caused by the biomass burning related emissions therein rather than the firework emissions observed during the DD phase.

In regard to the steep decline in ambient $PM_{2.5}$ following the fireworks event, similar observations were made by Hoyos et al., (2020), studying the impact of NYE fireworks on ambient air quality in Colombia. Hoyos et al., (2020) hypothesized that the steep, seemingly anomalous decline occurred due to atmospheric instability caused by a weak thermal inversion due to firecracker burning, leading to the rapid dispersion of aerosols and their episodic escape across the boundary layer. Hoyos and co-workers supported this hypothesis, using ceilometer-based backscattering intensity (BI) measurements, and observed anomalous BI readings marking the atmospheric instability following the fireworks event. Similar atmospheric behavior may be the reason behind the observations of the present study; however, the present study lacks any ceilometer-based or similar measurements to justify similar atmospheric behavior and leaves this question open for further investigation in the future. Another potential reason could be that firework emissions may have a limited impact in the regions upwind to the study site, and wind from these regions may aid in dispersing the pollutants to downwind areas. A future study may employ measurements similar to the present study in regions both upwind and downwind of Delhi to shed light on the role of meteorology in the steep decline in $PM_{2.5}$ following Diwali.

## 4. Conclusions

The present study reports near real-time variation of elemental compositions of $PM_{2.5}$ at two sites (IITMD and IITD) in New Delhi during 12 days around Diwali. The results indicate that the degree of enhancement in the concentration of species associated with fireworks displays is location-specific since considerable variability was observed between the two sampling sites. The ambient concentrations of metallic elements increased by a factor of ~22 to 320, depending on the particular species in the DD period compared to the ePD levels. The enhanced elemental concentrations pose a severe health hazard to the exposed population, since species like aluminum, barium, chlorine, lead, and manganese were noted to exceed the EPA risk-based concentration levels by orders of magnitude, at not only the peak levels but also the average levels during the DD phase. Moreover, potential carcinogens like As also exceeded the risk-based concentrations at both the IITD and the IITMD sites by order of magnitude.

The Diwali festival often coincides with the crop harvesting season, and the ambient air quality is seriously affected by increased stubble-burning events. These coinciding events often make it difficult to ascertain the impact of either of the two on ambient air pollution. The current study presents source apportionment results for highly time-resolved elemental and organic fractions of $PM_{2.5}$ to address that challenge. The $PM_{2.5}^{el}$ source apportionment was performed on the datasets from both the sampling sites separately, and both the analyses resulted in seven source profiles, i.e., industrial emissions, coal combustion, vehicular emissions, fireworks, dust-related emissions, biomass burning, and secondary chloride. The results of the source apportionment analysis lead us to conclude the following:

i) The fireworks impact is most prominent at both sampling sites only during the firecracker burning period. During the DD phase, the fireworks factor account for ~94% and ~92% of the total $PM_{2.5}^{el}$ at the IITMD and IITD sites, respectively; it is also important to note that the $PM_{2.5}^{el}$ levels at both locations were increased by a factor of ~12 in the DD phase as compared to the ePD levels.

ii) The coal combustion and dust-related sources mainly remained unaffected from the PD to the FD phase, whereas sources like industrial emissions, biomass burning, and vehicular emissions reduced by more than 50% at both the sites during the DD phase as compared to ePD mainly due to the reduced human mobility on the festival day. However, in the FD phase, industrial emissions and vehicular emissions trended back towards the ePD levels while remaining lower than the ePD average values, majorly due to 2-3 days of official holidays for the festivities.

iii) The biomass burning-related emissions rise steeply in the FD phase, and the average values increase by more than twice as compared to the ePD phase. In terms of the source apportionment of the organic fraction, the primary organic emissions display a distinct rise in concentrations in the DD phase. Also, both the POA and the two OOA factors display a considerable increase in the FD phase, hinting towards the role of biomass-related emissions towards the rise in primary organic emissions and, in turn, their aged products following the Diwali festival.

This is the first study to isolate the impact of fireworks on ambient $PM_{2.5}$ levels based on chemically speciated and high time resolution measurements of elemental $PM_{2.5}$ in Delhi. It is also the first study to delineate the potential impact of firecracker burning and the subsequent stubble burning events on the aerosol levels following Diwali night. These results pose important implications for guiding informed policies and control strategies to prevent/alleviate future haze events.

## Acknowledgments

This work was financially supported by the IRD Grand Challenge Project, IIT Delhi, under grant IITD/IRD/MI01810G, Department of Biotechnology (DBT), Government of India, under Grant BT/IN/UK/APHH/41/KB/2016-17 and by the Central Pollution Control Board (CPCB), Government of India, under Grant AQM/Source apportionment EPC Project/2017. The author, NH, would like to thank IIT Delhi for financial support through the Institutional Post-Doctoral Fellowship.

Table 1: Average and Maximum concentration levels of various inorganic species during the DD phase at the IITD and IITMD site, corresponding Generic Regional Screening Levels (RSLs) as suggested by US EPA and Potential Health Hazards

| Species | IITD Site (µg/m$^3$) | | IITMD Site (µg/m$^3$) | | US EPA Generic RSLs (µg/m$^3$) | | Potential Health Risk |
|---|---|---|---|---|---|---|---|
| | **Average** | **Maximum** | **Average** | **Maximum** | **Resident Air** | **Industrial Air** | **Carcinogenic / Non-Carcinogenic** |
| ALUMINIUM | 48.8146 | (114.8407) | 44.2154 | (178.3848) | 5.2 | 22 | None / Neurological |
| ARSENIC | 0.0105 | (0.0365) | 0.0172 | (0.0570) | 0.0007 | 0.0029 | Respiratory Cancer/Developmental |
| BARIUM | 14.3015 | (35.0304) | 15.4308 | (63.9083) | 0.52 | 2.2 | None / Cardiac |
| CHLORINE | 9.0537 | (17.7699) | 11.4043 | (34.9752) | 0.15 | 0.64 | None / Respiratory |
| LEAD | 0.6325 | (1.4244) | 0.8219 | (2.9588) | 0.15 | NA | None / Developmental |
| MANGANESE | 0.0735 | (0.1654) | 0.1004 | (0.4222) | 0.052 | 0.22 | None / Neurological |
| NICKEL | 0.0018 | (0.0046) | 0.0030 | (0.0114) | 0.012 | 0.051 | Lung Cancer/ Immunological |
| SELENIUM | 0.0004 | (0.0013) | 0.0005 | (0.0021) | 21 | 88 | None / Neurological |
| VANADIUM | 0.0238 | (0.0701) | 0.4305 | (3.8841) | 0.1 | 0.44 | None / Respiratory |

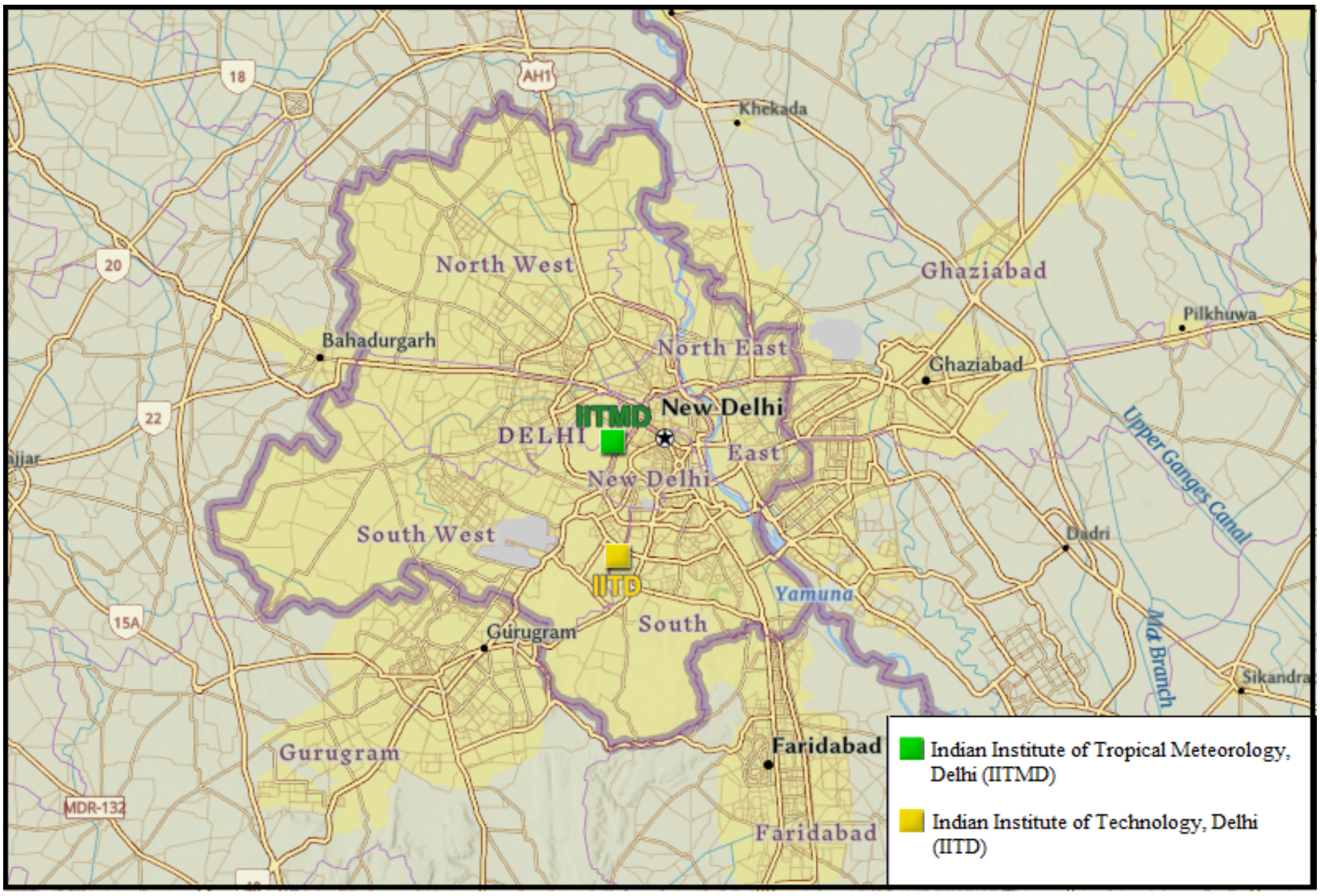

Figure 1: Location of the sampling sites in the present study: 1. Indian Institute of Tropical Meteorology, Delhi (IITMD) (green); 2. Indian Institute of Technology, Delhi (IITD) (yellow)

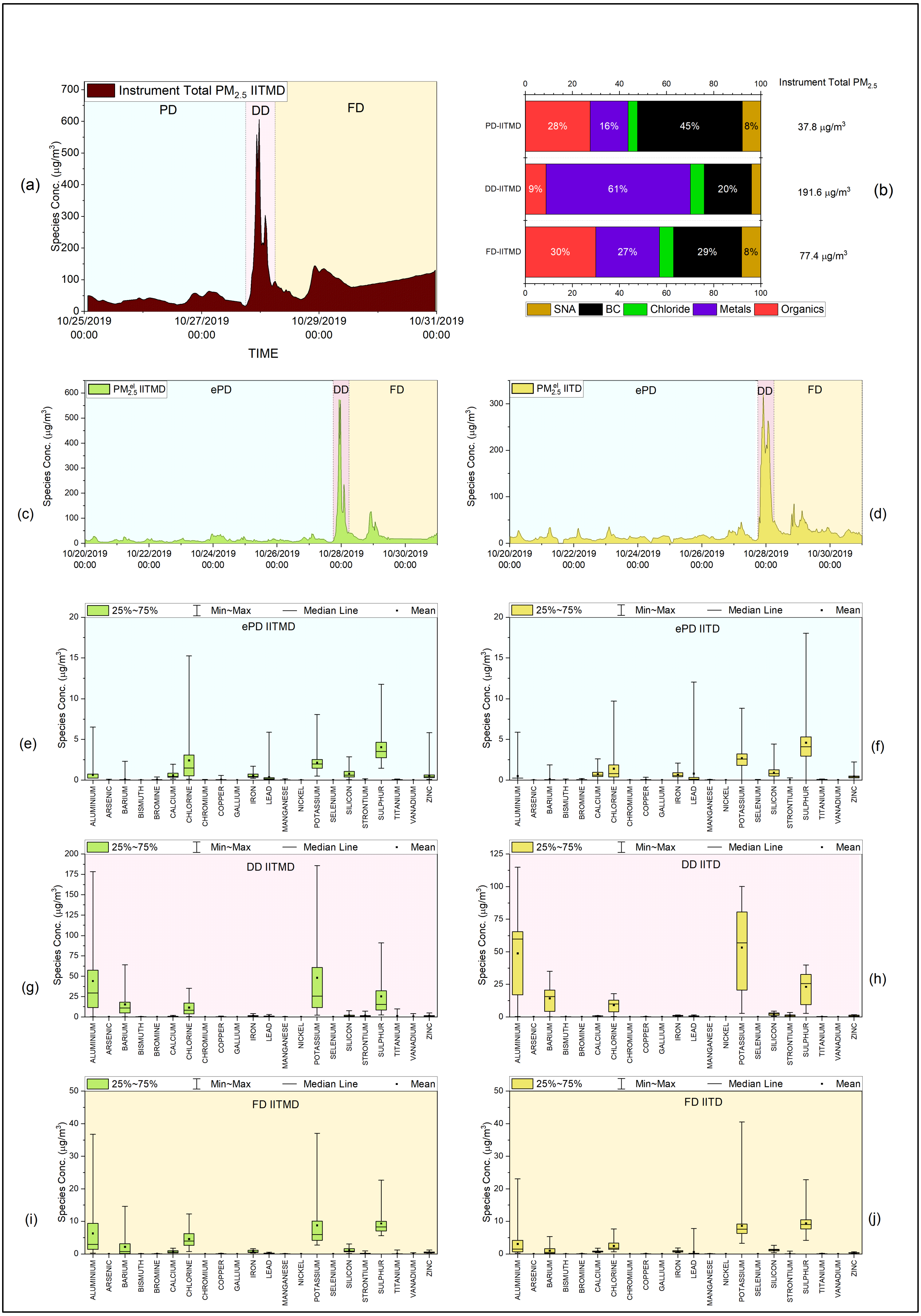

Figure 2: (a) Time variation of instrument total $PM_{2.5}$ at IITMD; (b) Variation in total $PM_{2.5}$ compositions in the PD, DD and FD phases; (c-d) Temporal Variation of $PM_{2.5}^{el}$ at the (left to right) IITMD (green) and IITD (yellow) sites

respectively; (e-j) Variation in constituents of $PM_{2.5}^{el}$ through each of (top to bottom) PD, DD and FD phase at the IITMD (green) and IITD (yellow) sites

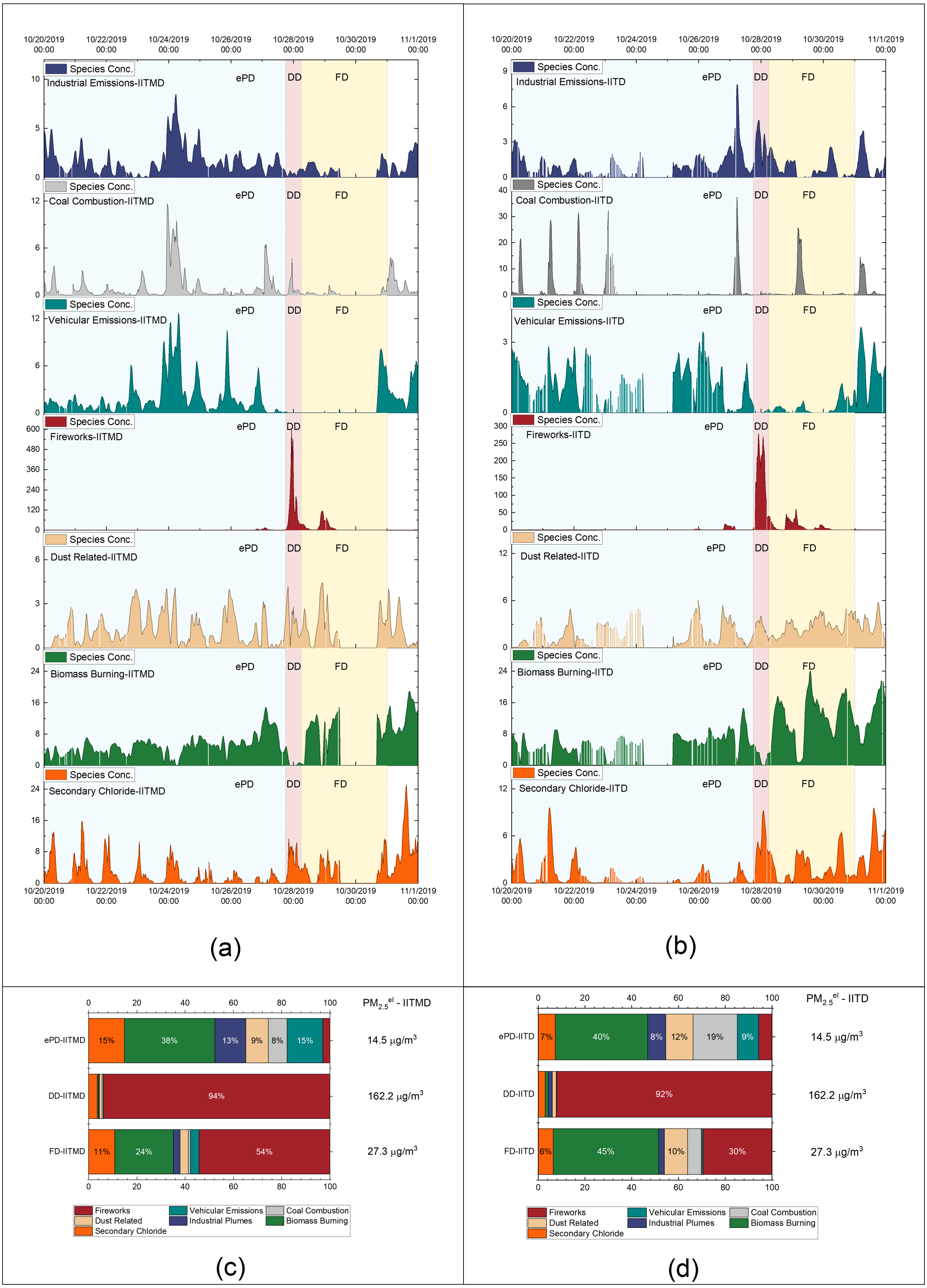

Figure 3: Temporal variation of PMF resolved sources of elemental $PM_{2.5}$; (top to bottom) industrial emissions, coal combustion, vehicular emissions, fireworks, dust-related, biomass burning, secondary chloride at (a) IITMD site; (b) IITD site; Phase-wise composition of elemental $PM_{2.5}$ at (c) IITMD site; (d) IITD site

# Supplementary Information for “Chemical speciation and source apportionment of ambient PM2.5 in New Delhi before, during, and after the Diwali fireworks.”

Chirag Manchanda [a], Mayank Kumar [a,*], Vikram Singh [b,*], Naba Hazarika [c], Mohd Faisal [b], Vipul Lalchandani [d], Ashutosh Shukla [d], Jay Dave [e], Neeraj Rastogi [e], Sachchida Nand Tripathi [d,*]

a. Department of Mechanical Engineering, Indian Institute of Technology Delhi, New Delhi, India

b. Department of Chemical Engineering, Indian Institute of Technology Delhi, New Delhi, India

c. Department of Applied Mechanics, Indian Institute of Technology Delhi, New Delhi, India

d. Department of Civil Engineering, Indian Institute of Technology Kanpur, Uttar Pradesh, India

e. Geosciences Division, Physical Research Laboratory, Ahmedabad, India

## S1. Quality Assurance and Quality Control (QA/QC) Procedures

For the duration of the study period, the Xact 625i at both sites operated at a half-hourly time resolution, sampling ambient $PM_{2.5}$ at the standard airflow rate of 16.7 lpm. Automated quality assurance checks were performed by the instrument every midnight for the elements Cr, Cd, and Pd. Other than the automated checks, XRF and flow calibration, along with leak and flow checks, were performed as per the maintenance routine advised by the manufacturer.

NR-$PM_{2.5}$ was analyzed using an HR-ToF-AMS operating at a standard flow rate of around 0.1 lpm, while sampling and analyzing at a 2-min interval, with background air allowing for continuous air beam correction. The collection efficiency (CE) was determined for the mass spectra for an optimum collection considering inlet humidity, particle composition, and aerosol acidity, as described by (Middlebrook et al., 2012). A monodisperse supply of $NH_4NO_3$ and $(NH_4)_2SO_4$ aerosols, selected through a differential mobility analyzer (DMA) and counted using a condensation particle counter (CPC), was used to calculate the ionization efficiency (IE) of $NO_3$ and relative ionization efficiencies (RIE) of $NH_4$ and $SO_4$ as suggested by (Crenn et al., 2015).

Near real-time concentrations of Black Carbon were recorded using a multichannel Aethalometer operated at a unit time resolution and a 5 lpm ambient air flow rate. The readings were monitored continuously for any instrument warnings, noise/spikes, as well as BC vs. attenuation values for any required compensation. Other than that, regular flow rate checks, inspection, and cleaning of the optical chamber and insect screen assembly were conducted as a part of routine maintenance.

## S2. Positive Matrix Factorization

### *S2.1. PMF algorithm*

As discussed in section 2.1, Positive Matrix Factorization (PMF) is a statistical tool enforcing chemical mass balance utilizing receptor-based measurements. The PMF algorithm implements non-negativity constraints on each of the solution matrices to prevent negative factor concentrations. The elements constituting the matrices **G** and **F** are iterated to minimize the objective function or the Q-value given by eq. (1):

$$Q = \sum_i \sum_j \left(\frac{e_{ij}}{s_{ij}}\right)^2 \quad ..... (1)$$

* Corresponding Authors
E-mail addresses : kmayank@mech.iitd.ac.in (M. Kumar); vs225@chemical@iitd.ac.in (V. Singh); snt@iitk.ac.in (S.N. Tripathi)

Here, the term $s_{ij}$ represents the uncertainty associated with each element $x_{ij}$ of the input data matrix **X**. Thus, the PMF algorithm allows each input data point to be weighted individually, enabling the user to alter the impact of each data point on the final solution, depending on the confidence in the associated measurement. Hence, low confidence measurements, like for the case of concentrations below method detection limits, can be down-weighted w.r.t the other measurements, i.e., the ones above MDL.

*S2.2 PMF input preparation*

The present study employed the Xact 625i for measuring $PM_{2.5}^{el}$ at both the sampling sites, i.e., IITMD and IITD. Both the instruments were configured to measure 43 elements in total (Al, Si, P, S, Cl, K, Ca, Sc, Ti, V, Cr, Mn, Fe, Co, Ni, Cu, Zn, Ga, Ge, As, Se, Br, Rb, Sr, Y, Mo, Pd, Ag, Cd, In, Sn, Sb, Te, Cs, Ba, La, Ce, Pt, Au, Hg, Tl, Pb, and Bi) with a half-hourly time resolution. For the duration of the study, the concentration levels of 18 elements (P, Sc, Co, Ge, Y, Mo, Ag, Cd, In, Sn, Te, Cs, La, Ce, Pt, Au, Hg, Tl) were found to remain below the Minimum Detection Limit (MDL) for more than 50% of the readings and were thus ignored for the PMF analysis (Polissar et al., 1998). The uncertainty $s_{ij}$, associated with every data point is reported by the ADAPT software module, on-board the Xact 625i. The reported uncertainty accounts for the measurement uncertainty, as well as the spectral deconvolution uncertainty (Tremper et al., 2018), and the same has been employed for preparing the PMF input uncertainty matrix in the present study. The datasets from both instruments were subjected to PMF individually.

The non-refractory fraction of $PM_{2.5}$ (NR-$PM_{2.5}$) was studied using the HR-ToF-AMS, which determines the quantitative mass spectra (MS) of NR-$PM_{2.5}$ up to a mass to charge (*m/z*) ratio of 200. The measured MS is further resolved into the sulfate, nitrate, ammonium, chloride, and organic fractions using a library of pre-determined fragmentation characteristics, as employed by (Allan et al., 2004). The uncertainty values $s_{ij}$ associated with the species measured using the AMS were calculated using the standard HR-ToF-AMS data access software, following the procedure documented by (Ulbrich et al., 2009) and (Ng et al., 2011a).

The MDL values for each elemental species measured using the Xact were provided by the manufacturer, i.e., Cooper Environmental Services (CES), while the MDLs for the organic MS measured using the HR-ToF-AMS were determined according to (Drewnick et al., 2009). The missing data values caused by power failure and maintenance shutdown were neglected from the PMF input (Rai et al., 2020).

The input measurements corresponding to each species $j$ were screened further on the basis of the signal-to-noise ratio (S/N) computed within the EPA PMF 5.0 software module (Norris et al., 2014). For each entry $x_{ij}$ in the input data matrix and its corresponding uncertainty $s_{ij}$, the difference between the two is regarded as the signal, such that

$$d_{ij} = \left(\frac{x_{ij} - s_{ij}}{s_{ij}}\right) \quad \text{if } x_{ij} > s_{ij} \quad ...... (2)$$

$$d_{ij} = 0 \quad \text{if } x_{ij} < s_{ij} \quad ...... (3)$$

The S/N ratio is then given by the following equation:

$$\left(\frac{S}{N}\right)_j = {}^{1}\!/_{n} \sum_{1}^{n} d_{ij} \ldots\ldots (4)$$

Based on the S/N value, each input species is further characterized as strong, weak, or bad (Norris et al., 2014). The species with S/N > 1 were categorized as strong, and input concentration and uncertainty were used for further analysis with no alteration. The species with the S/N ratio between 0.5 and 1 were categorized as weak; for such a case, the input uncertainty values were augmented by a factor of four and used along with the input concentration for further analysis. Finally, the species with an S/N ratio of less than 0.5 were classified as bad signals and were excluded from further analysis.

In the present study, in the case of the Xact-based PMF, Rb, Sb, and Pd were identified as bad signals and were ignored from the PMF input. However, none of the *m/z* signals measured using the HR-ToF-AMS fell into the category of bad signals.

*S2.3. Factor Selection and Uncertainty Quantification*

The quality of source profiles resolved using the PMF analysis often relies on the number of factors deemed optimal, for representing the input dataset, by the user. However, the PMF algorithm is focused on minimizing the sum of weighted residual errors for a linear fitting problem. A singular solution is seldom possible for such a scenario as there can be several possible solutions with similar residuals; this led multiple past studies to conclude that mathematical diagnostics alone are insufficient for deciding on the right number of factors (Canonaco et al., 2013; Rai et al., 2020). Thus, the present study explores the physical realizability of each resolved factor by comparing its temporal variation with recognized external tracers, evaluating for observed diurnal behavior, known elemental or *m/z* dominance in the factor profile, in addition to the established diagnostics like Q-value, $Q/Q_{exp}$ (Canonaco et al., 2013) and scaled residuals for each species.

The solution space corresponding to the input measurements is assessed with a set of initial base runs, wherein the number of factors is varied from 2 to 10 with 10 seeds each (number of PMF runs with pseudorandom starting points). The same procedure is adopted for all 3 datasets analyzed in the present study (2 Xact-based + 1 AMS-based). The vicinity of a plausible solution, i.e., the optimal number of factors resulting in the lowest Q-value and no significant change in Q-value by further increasing the number of factors, is further explored by repeating the PMF runs with 50 seeds each.

The solution corresponding to the number of factors deemed optimal after the base run is tested for rotational ambiguity using the DISP analysis available within the EPA PMF 5.0 module. The DISP technique assesses the maximum possible variation in the source profile obtained as the results of the base run, without an appreciable change in the base Q-value. In EPA PMF 5.0, the DISP tool perturbs each species, one at a time, in the source profile resolved from the optimal base run, and following each alteration, the PMF run is repeated to evaluate changes in the source profile due to each perturbation, such that the change in the Q-value w.r.t the base run remains under a pre-determined maximum change $dQ_{max}$ ($dQ_{max}$ = 4,8,15,25). These perturbations, at times, may lead to a switch in the identity of the factor profile w.r.t the base run, i.e., a particular base factor after displacement may be better correlated with another base factor than the original base profile, such a case is noted to be a factor swap. These swaps are accounted for using uncentered cross-correlations between the original and the displaced factor profiles

No factor swaps were observed in the present study for both the Xact and AMS-based datasets. In the present study, the maximum decrease in Q-value for the IITMD Xact dataset was observed to be 0.027 with % dQ change as 0.0021%; in the case of the IITD Xact dataset, the maximum decrease in Q-value was found to be 0.038 with % dQ change as 0.0039%, while for the IITMD HR-ToF-AMS dataset the largest decrease in Q-value was observed to be 1.486 with % dQ change as 0.00363%. Past studies have established % dQ change within 1% to represent acceptable DISP solutions (Norris et al., 2014).

The optimal base-run solutions were further subjected to controlled rotations via variation in f-peak values from -1 to +1 with an increment of 0.1 to evaluate the impact of rotations on parameters like the fraction of variance explained by each factor, the mutual correlation between the time series of resolved factors, correlation of factor profile with reference profiles and correlation of factor time series with external tracers (Bhandari et al., 2020; Rai et al., 2020; Ulbrich et al., 2009). The range of the f-peak value was limited to ensure that the % change in Q-value between the rotated and base factor profile is small. In the case of the IITMD and IITD Xact based datasets, a minimal effect of factor rotation was observed in terms of the mutual correlation between the factors and correlation with external tracers; thus, in both cases, the base solution was considered optimal. However, for the IITMD HR-ToF-AMS dataset, f-peak variations were found to reduce mutual correlations between the factor time series and improve correlations between factor profiles and reference MS from (Ng et al., 2011b). The solution corresponding to the fpeak value of +0.2 was found to best improve the correlations with minimal change in Q-value, thus was chosen as the optimal solution (Bhandari et al., 2020; Rai et al., 2020).

Further, the bootstrap (BS) randomized resampling strategy was employed to evaluate the effect of random error on the solution (Brown et al., 2015). Similar to the DISP analysis, the BS analysis is effected through the EPA PMF module; it works by creating a randomized input matrix of size equal to the original input matrix, using non-overlapping blocks of input measurements, where the user determines the block size. The new input matrix is subjected to PMF, and the BS factors are mapped back to the base factors based on the uncentered correlations exceeding a user-defined threshold. In case a BS factor fails to have an uncentered correlation with any of the base factors greater than the threshold, then that BS factor is deemed unmapped. The BS results act as a proxy to understand the uncertainty associated with the resolved factors and the species associated with each factor profile. In the present study, the optimal base/f-peak solutions were subjected to 800 BS runs, with an uncentered correlation threshold of 0.8. In both Xact based and AMS-based datasets, none of the BS factors remained unmapped. In the case of the IITMD Xact dataset, 738 BS runs, for the IITD Xact dataset, 721 runs, and for the IITMD HR-ToF-AMS dataset, 742 BS runs were deemed reasonable solutions, i.e., the BS factor was mapped back to the original base factor.

As discussed in section 3.2, to ascertain whether the peaks associated with the fireworks in the ePD and FD phases are due to scattered firecracker burning events before and after Diwali day or just statistical remnants from the PMF analysis, we attempted to constraint the firework emissions to the DD phase. The imposed constraints were found to impact the source profiles and variation of other sources. When this solution was subjected to the BS analysis, only 569 of the 800 BS runs were found to represent reasonable solutions indicating that the initial solution is significantly better in encompassing the observed variation compared to the constrained solution. Increasing number of factors leads to further degradation in number of acceptable BS runs for both the constrained and original solution indicating overfitting in the case of more factors.

## S3. Supplementary Results

### *S3.1 Tracers marking firework events*

As discussed in section 3.1 of the main manuscript, elements namely K, Al, Sr, Ba, S, and Bi were identified as elemental tracers to mark the fireworks event in the case of the present study. This section elucidates the connection of these tracers to firecracker constituents.

Multiple studies around the world have recognized potassium as a reliable marker for particulate emissions from firecracker burning events (Shon et al., 2015; Tanda et al., 2019; Tian et al., 2014; Yang et al., 2014). K is known to be a significant component of the black powder fuel, which is a common raw material in most fireworks and is present in the form of $KClO_3$ or $KNO_3$ (Drewnick et al., 2006; Dutcher et al., 1999). However, in India, the period of October – November, during which Diwali usually occurs, witnesses a rise in the stubble and biomass burning events (Jethva et al., 2019) also, as it coincides with the crop harvesting season and the onset of winters. Potassium is also widely accepted as a tracer for the biomass and the stubble burning events (Pant and Harrison, 2012); thus, such activities may impose background disturbances when considering K as a sole marker in the context of Diwali.

Similar to K, nitrate and perchlorate salts of Ba are used as oxidizers in fireworks to facilitate rapid combustion (Do et al., 2012). Metals like Al, Ba, and Sr are often used as coloring agents for fireworks; Al salts have been known to emit white and silver sparkles, while Ba gives green as chlorates, white as carbonates and nitrates, while Sr salts are known to emit red (Kulshrestha et al., 2004; QW et al., 2018). Bi compounds ($Bi_2O_3$, $(BiO)_2CO_3$) are mostly used to produce crackling stars in fireworks (Perrino et al., 2011; Tanda et al., 2019). Sulfur, similar to potassium, is another component of the black powder fuel used in firecrackers and is used as a reducing agent (Drewnick et al., 2006; Dutcher et al., 1999).

As discussed in section 3.1, in addition to the elemental tracers, the BC levels and the $SO_4$ to $NO_3$ ratio were also found to display a distinct peak during the DD phase, thus, acting as a marker to the firework displays. The BC emissions from firecrackers are usually traced back to the combustion of the carbonaceous matter, which acts as a reducing agent in the black powder fuel (Babu and Moorthy, 2001; Dutcher et al., 1999).In the case of the $SO_4$ to $NO_3$ ratio, the potential reason for the enhanced levels, as presented by past studies, rests on the hypothesis that $SO_4$ is directly emitted from firecracker burning due to oxidation of sulfur present in the black powder fuel, thus supplementing the sulfate from $SO_2$ oxidation from the atmosphere. At the same time, nitrate continues to remain in the photo-stationary state with $O_3$ and $NO_x$ for oxidation (Feng et al., 2012).

### *S3.2 Potential health hazards associated with firework emissions*

It has been well accepted that increased exposure to respirable particulate matter can have a deleterious impact on human health (WHO, 2013). Past studies have also noted that the elderly and children and those with a past history of heart or lung diseases are more susceptible to the pernicious health effects (Schwartz et al., 1996). It comes as a natural consequence that enhanced $PM_{2.5}$ concentrations observed during Diwali (Figure 2(a)) lead to an increased health risk for the exposed population. However, as noted earlier, in the case of Diwali, the rise in the $PM_{2.5}$ levels stems from the escalated concentrations of certain elemental species (Figure 2 (e-f) & Figure S3) originating from firework displays. It is essential to take note that most of these elemental species are toxic in nature and pose a severe health risk to the exposed population, ranging from cardiopulmonary and respiratory disorders to neurological impact. Inorganic species like Al, Sr, Ba, Bi, and BC were identified as tracers to mark the fireworks event in the present study (Figure S3). Past studies have noted the potential role of increased exposure to aluminum in the development of Alzheimer's disease and other neurological disorders (Niu, 2018), while exposure to increased levels of strontium has been noted to impact respiratory and bone health (Agency for Toxic Substances and Disease Registry, 1999). Exposure to barium salts have associated with neurological, musculoskeletal, and respiratory effects (Effiong and Neitzel, 2016), prolonged exposure to bismuth compounds have been recognized to lead to neurotoxicity and nephrotoxicity (Wang et al., 2019); past toxicological studies have also associated BC exposure with cardiopulmonary morbidity (Janssen et al., 2013).

### *S3.3 Firework emission levels during Diwali compared to similar events*

As discussed in the main manuscript, events involving firework displays are common to almost all parts of the globe; however, it is essential to take note of the levels of emissions associated with Diwali w.r.t similar events across different parts of the world. In a recent study, (Tanda et al., 2019) studied the impact of firework displays during the New Years' Eve (NYE) on PM (ranging from 15nm to 10 μm) at Brno, Czech Republic, and Graz, Austria. At Brno, the average concentrations during the NYE for each of Al, Ba, K, Sr, and Bi were reported to be 54, 23, 381, 6.1, and 2.8 ng/m$^{3,}$ respectively. Similarly, at Graz, average concentrations for each of Al, Ba, K, Sr, and Bi were found to be 29, 14, 460, 4.8, and 1.2 ng/m$^{3,}$ respectively. Thus, while the study by (Tanda et al., 2019), accounts for heavier particles ( >2.5 μm) as compared to the present study, the average concentration of trace metals remain 2-3 orders of magnitude lower than the present study, when compared to not only the DD phase but also the ePD phase. Similar conclusions have been reached upon by an extensive review of such studies conducted by (Lin, 2016). While the degree of enhancement in concentrations and the average concentration levels of individual metals recorded by us are similar to those reported by some past studies focusing on some of these metals during Diwali across different parts of India (Chatterjee et al., 2013; Perrino et al., 2011; Sarkar et al., 2010), it is essential to take note that none of these studies have been able to provide a holistic view of a variation of $PM_{2.5}{}^{el}$ at a time resolution and level of chemical speciation as in the present study, majorly due to the limitations associated with filter-based techniques employed by these studies. However, a few studies in different parts of the world have employed high-time resolution equipment similar to ours to quantify the impact of firework events, enabling them to comment on total $PM_{2.5}{}^{el}$. (QW et al., 2018), focused on the Spring festival in Chengdu, China, to quantify the impact of fireworks on $PM_{2.5}{}^{el}$; however, the average $PM_{2.5}{}^{el}$ concentration recorded during the fireworks event was recorded to be 24.1 μg/m$^3$, which is still much lower than the average levels in the present study (155 μg/m$^3$), further highlighting the intensity of emissions associated with the firecracker burning associated with the Diwali event.

### *S3.4 Sources of elemental $PM_{2.5}$*

As discussed in section 3.2, the sources of elemental $PM_{2.5}$ were resolved using PMF. The real-world sources associated with each of the resolved factors were determined on the basis of the dominant species corresponding to each resolved factor. The species contribution in each factor was quantified in terms of both the percentage of factor total (% FT) (concentration of species of interest as a fraction of total factor concentration) and the percentage of species sum (%SS) (ratio of the concentration of species of interest in the factor of interest to the total species concentration across all factors). The details on the composition and relevance of each resolved factor to real-world sources are as follows:

*S3.4.1 Industrial Emissions*

At the IITMD site, the particulate matter apportioned to the industrial plumes (figure S4(a)) is dominated by S (57% of FT), followed by Zn (26% of FT) and K (10% of FT). In terms of % SS, the industrial plumes contribute to 66% of Zn, 29% of each Br and V, 18% of Se, and 16% of Cr, while around 13% of total S. At the IITD site, the resolved factor (figure S4(b)) is dominated by S (44% of FT), followed by K (18% of FT) and Zn (10% of FT). In terms of % SS, the factor at IITD is found to contribute to 48% of total Cu, 44% of As, and 25 % of Zn.

While both these independently calculated profiles present some disparities in terms of the % SS that the factor accounts for, it is essential to note that both factors were found to be dominated by Zn, K, and S in terms of % FT. The dissimilarities seem to be indicative of effects localized to each receptor site, like local wind direction or some localized industrial sources specific to each location, leading to the disparities.

Some past studies have noted elements like Zn, K, As, and V to be makers for iron and steel industries (Duan and Tan, 2013; Vincent and Passant, 2013). A study by (Negi et al., 1987) differentiated between non-ferrous industrial emissions( Zn, Cu, Mn), textile emissions (V, Br), and oil refinery emissions (S, Cu, V). (Shridhar et al., 2010) characterize Zn, Cu, Cr to originate from industrial emissions, while (Mouli et al., 2006) traced Se to originate from metallurgical industries. The overlap and the ambiguity between the tracers proposed by various past studies to mark different kinds of industries have led us to define the factor to be associated with industrial emissions rather than attempting to characterize it to the type of industry further.

*S3.4.2 Coal Combustion*

At the IITMD site (figure S4 (a)), the resolved coal combustion source profile is dominated by S (34% of FT) followed by Pb (28% of FT), while in terms of %SS, coal combustion accounts for 71% of total Pb, 41% of As, 22% Cu and 15% of total Se. Again at the IITD site (figure S4 (b)), in terms of %FT, the factor is dominated by S (44% of FT), followed by Pb (28% of FT), while in terms of %SS, it contributes to 82% Pb and 60% Se.

Se and As are well accepted as tracers for coal combustion by many studies both in India and across the world (Hien et al., 2001; Lee et al., 2008; Sharma et al., 2007). While commercially available coal is limited in Pb-content, past studies have noted Pb levels to be higher in domestic coals (Negi et al., 1987). However, local coal in northern India is known to have low sulfur content due to aged deposits; the high sulfur levels in the resolved factors may indicate the use of sulfur-rich coal from northeastern India (Chandra and Chandra, 2004; Sarkar, 2009). However, it is also known that power plants utilize blends of domestic and imported coal to get a balance between low ash content and high sulfur content (Central Electricity Authority, 2012; Chandra and Chandra, 2004), signaling the possible reception of coal-burning associated aerosols from thermal power plants. Again, the real-world source to such a factor profile may also correspond to small-scale industries. Some studies have reported Pb aerosol levels to be enhanced during lead smelting which may mix with coal combustion related particulate matter from the same plants to result in the profile resolved in this study.

*S3.4.3 Vehicular Emissions*

The vehicular emissions factor at the IITMD site (figure S4 (a)) is dominated by K (24% of FT), followed by Cl (17% of FT), S (16% of FT), Si (12% of FT), and Fe (10% of FT), whereas in terms of the %SS the vehicular emissions factor accounts for 76% of total Cr, 61% of Mn, 55% of Ni, 43% of Ti, 27% of Fe, 23% of Si and 22% of Ca. At the IITD site (figure S4 (b)), the vehicular emissions are dominated by Ca (19% of FT), Si and Fe (18% each by FT), S (16% of FT), and Zn (15% of FT), looking at the %SS this factor contributes to 83% of total Cr, 48% of Ni, 32% of Zn, 23% of Mn, 22% of Ca and 21% of Fe.

Potassium is known to be used as an additive in engine oil as well as an anti-freeze inhibitor. In addition to that, K is also recognized to be present in almost all unleaded fuels (Spencer et al., 2006). Chlorine-based lubricating oils are used in the form of dispersants, which are used to protect the engine by retaining dirt in suspension (Dyke et al., 2007). Sulfur is known to occur naturally in crude oil. As per the currently applicable emission standards in India, i.e., BS-IV, diesel fuels are permitted to contain up to 50 ppm of sulfur (Transport Policy, 2015). Also, some past studies have pointed out the use of sulfur in engine oil anti-wear additives (Fitch, 2019). Calcium is known to be added to engine lubricants to serve as a base to neutralize acids and to stabilize highly polar compounds formed in the engine as combustion bi-products (Lyyränen et al., 1999; Rudnick, 2017).

Past studies have attributed Fe, Mn, and Zn to brake and engine wear emissions, citing their abundance in brake lining and brake pads (Gianini et al., 2012; Grigoratos and Martini, 2015; Thorpe and Harrison, 2008). Some studies have also attributed Ti and V to brake and tire wear (Gerlofs-Nijland et al., 2019). Ni and Cr emissions have been associated with Ni-Cr-based catalytic converters employed for controlling toxic vehicular emissions (Negi et al., 1987; Srivastava et al., 2016).

*S3.4.4 Fireworks*

At the IITMD site (figure S4 (a)), the fireworks source profile is dominated by K (31% of FT), Al (30% of FT), and S (20% of FT), while based on the %SS, this factor contributes about 91% of Ba, 87% of Bi, 86% of Sr, 80% of Al, 59% of Ga, 54% of K and 43% of V. At the IITD site the factor profile (figure S4 (b)) is again dominated by K (37% of FT), followed by Al (30% of FT) and S (20% of FT), in terms of the % SS the fireworks contribute to 87% of Ba, 85% of Bi, 85% of Al, 83% of Sr, 56% of V, 55% of Ga and 48% of K.

Most of the dominating species noted above, including K, Al, S, Ba, Bi, and Sr, have already been recognized as constituents of firecrackers and, in turn, tracers for marking firework events and have been discussed in detail in section S3.2. However, Ga and V, as discussed in section 3.1, remained near MDL for most of the non-Diwali period and raised to considerable concentrations only in the DD phase. However, limited studies in the past have observed an increment in Ga and V concentrations during a fireworks event (Dutcher et al., 1999; Perrino et al., 2011).

*S3.4.5 Dust-Related*

The dust-related factor at the IITMD site (figure S4 (a)) is dominated by Si (26% of FT), Ca (24% of FT), Al (16% of FT), and Fe (12% of FT), while in terms of %SS, the factor accounts for 46% of total Ca, 34% of Si, 24% of Fe, 23% of V. At the IITD site the dust-related factor (figure S4 (b)) is dominated by Si (34% of FT), Ca (26% of FT), and Fe (24% of FT). In contrast, in terms of % SS the dust-related factor contributes around 70% of total Ca, 64% each of Si and Fe, 61% of Mn, 49% of Ti, 35% of Ni, and 36% of Cu. It is important to note that while the dust-related factor at IITD seems to account for 50% of total measured Ti, the average total Ti concentrations are of the order of ~0.05 $\mu g/m^3$.

Multiple studies across the globe have used crustal elements including Ca, Al, Si, Ti, and Fe, as tracers to soils dust or crustal re-suspension (Jaeckels et al., 2007; Kothai et al., 2008; Lough et al., 2005; Rai et al., 2020; Stone et al., 2010). However, elements like Ni, Cu, and V may have contaminated the crustal dust over time, with actual origins

*S3.4.6 Biomass Burning*

At the IITMD site (figure S4 (a)), this factor is dominated by S(51% of FT) and K (25% of FT), while in terms of %SS, the factor contributes to around 66% of Se, 51% of Ti, 46% of S, 43% of V and 32% of Br. At the IITD site, the biomass burning source profile (figure S4 (b)) is dominated by S (50% of FT) and K (34% of FT), whereas in terms of %SS, it accounts for 58% of Br, 55% each of S and As, 39% of Se and 37% of K.

Potassium has been a widely accepted marker for biomass burning by several past studies (Khare and Baruah, 2010; Pant and Harrison, 2012; Reche et al., 2012; Shridhar et al., 2010). A study by (Li et al., 2003) concluded that for fresh biomass burning plumes, much of the potassium content exists as KCl or, to a degree KBr, while in the case of aged plumes, the chloride and bromides are partially substituted by sulfates, thus supporting the resolved profile in our case for high S and Cl concentrations in conjunction with K in the biomass burning source profile. Prior studies have found selenium to reach significant levels in biomass grown in Se-rich soils (Goldstein, 2018), which in turn are typical to northern India (Sharma et al., 2009). Observations similar to Se have been made for Ti, for enrichment through the soil; however, in relatively lesser concentrations (Tlustoš et al., 2011). Pyrolysis of fibers in biomass, like cereal, grass, and straw, has been noted to emit silicon (Obernberger et al., 2006). Toxic elements like V and As while representing considerable quantities in terms of % SS, but the actual concentration associated with the factor profiles are of the order of $10^{-4}$ μg/m$^3$ and $10^{-3}$ μg/m$^3$ respectively, such traces concentration may result from some amount of anthropogenic contamination, similar toxic metal concentrations in biomass emissions have been noted by past studies (DEMİRBAŞ, 2003).

*S3.4.7 Secondary Chloride*

The secondary chloride factor is solely dominated by Cl, accounting for 78% and 72% of FT at the IITMD and IITD sampling sites (figure S4), respectively. In terms of the %SS, the source profile at the IITMD site contributes to 62 % of total Cl and 32% of total Br, while the factor resolved at IITD accounts for 53% of total Cl and 26% of total Br.

***S3.5. Source apportionment of organic $PM_{2.5}$***

The organic fraction of total $PM_{2.5}$ measured using the AMS at the IITMD site was subjected to source apportionment using PMF analysis. The variation in the mass spectra was best described by a set of 3 factors, i.e., one primary organic aerosol (POA) and two oxygenated organic aerosols (OOA) profiles. These results are similar to the results obtained by (Bhandari et al., 2020), wherein a POA and an OOA factor were resolved during the winters of 2017 and 2018. The resolved factors were identified based on the resolved mass spectra, which were further characterized using AMS reference profiles from (Aiken et al., 2009) and (Ng et al., 2011b). The detailed description for each of the resolved factor profiles is as follows:

*S3.5.1 Primary Organic Aerosol (POA)*

The POA mass spectra (figure S5(a)) is marked by distinct peaks of *m/z 55* and *m/z 57,* which are characteristic of primary organic emissions like traffic and cooking-related emissions (Zhang et al., 2005; Zhu et al., 2018). The resolved mass spectra have a significant correlation with the hydrocarbon-like organic aerosol (HOA) reference profiles from both (Aiken et al., 2009) (Pearson R = 0.92) and (Ng et al., 2011b) (Pearson R = 0.89). Past studies have often employed BC as a tracer for POA emissions (Schmidt and Noack, 2000; Zhu et al., 2018), as BC is mainly derived from incomplete combustion products. In the present study, a significant correlation (Pearson R = 0.89) between BC and POA was observed (figure S5 (b)). The diurnal characteristics (figure S8 (a)) present variation characteristic to POA (Zhu et al., 2018), with a strong dependence on BLH, resulting in a distinct peak around 24:00 IST with reducing intensity with increasing in BLH towards the morning.

In terms of the time variation of this factor presented in figure S5(b) as well as the change in percentage contribution to total organic content (figure S5(c)), we observe a steep rise in POA concentrations during the DD phase, indicating that the POA factor captures the direct organic emissions from the firecracker burning during Diwali. The average factor concentrations during the DD phase are around 7.8 times the PD levels; however, the concentrations remain around 5 times higher than the average PD levels in the FD phase. This behavior may arise from the primary organic emissions associated with the increasing biomass burning concentrations towards the FD phase as observed in section 3.2.6 during the $PM_{2.5}^{el}$ source apportionment; the appreciable correlation observed between POA and BC bolsters this hypothesis further as biomass burning is known to contribute significantly to total BC (Pant and Harrison, 2012), and BC-like POA continues to increase in the FD phase.

*S3.5.2 Oxygenated Organic Aerosols (OOA-1 and OOA-2)*

The OOA-1 and OOA-2 mass spectra are marked by distinct peaks of *m/z 44* or $CO_2^{+,}$ which are tracer fragments used to mark Oxygenated Organic Aerosol (OOA) (Aiken et al., 2009; Li et al., 2019). The mass spectra associated with both OOA-1 & 2 display significant correlation with reference OOA profiles from (Aiken et al., 2009) (Pearson R = 0.92, 0.95 for OOA-1, OOA-2 respectively) and also reference LVOOA profiles from (Ng et al., 2011b) (Pearson R = 0.80, 0.84 for OOA-1, OOA-2 respectively). Both the factors show diurnal profiles (figure S8 (b & c)), with a sharp peak in concentration during noon, which is characteristic of OOA diurnal behavior (Aiken et al., 2009), due to enhanced photochemical aging of POA during noon.

In terms of the time variation of the OOA factors presented in figure S5(b) as well as the change in percentage contribution to total organic content figure S5(c), we observe a slight increment in the OOA concentrations during the DD phase, suggesting a slight immediate impact of the fireworks on the OOA concentrations. OOA-1 concentrations during the DD phase rose by around 30%, while OOA-2 concentrations rose by around 22% as compared to the PD averages. However, we note that there is a steep rise in the OOA concentrations in the FD phase, while OOA-1 rises by around 50% compared to its PD levels. The average OOA-2 in the FD phase rises about 186% as compared to the PD phase. While the Diwali event may have some contribution to this rise in OOA levels due to oxidation of the organic aerosols emitted on the Diwali night, however, the increased OOA levels may also indicate an increased influx of aged aerosol from biomass burning activities rising during the period affecting both organic and inorganic fractions of total $PM_{2.5}$.

The steep rise in OOA-1 and OOA-2 following Diwali, along with increasing elemental biomass burning contributions as noted in section 3.2.6, together account for the increasing background $PM_{2.5}$ levels, observed in Figure 2 (a-b), following Diwali.

Figure S1: Linear regression analysis comparing Instrument Total $PM_{2.5}$ (Xact 625i + HR ToF AMS + Aethalometer AE33) (y-axis) with Beta Attenuation Monitor (BAM) measured $PM_{2.5}$ (x-axis)

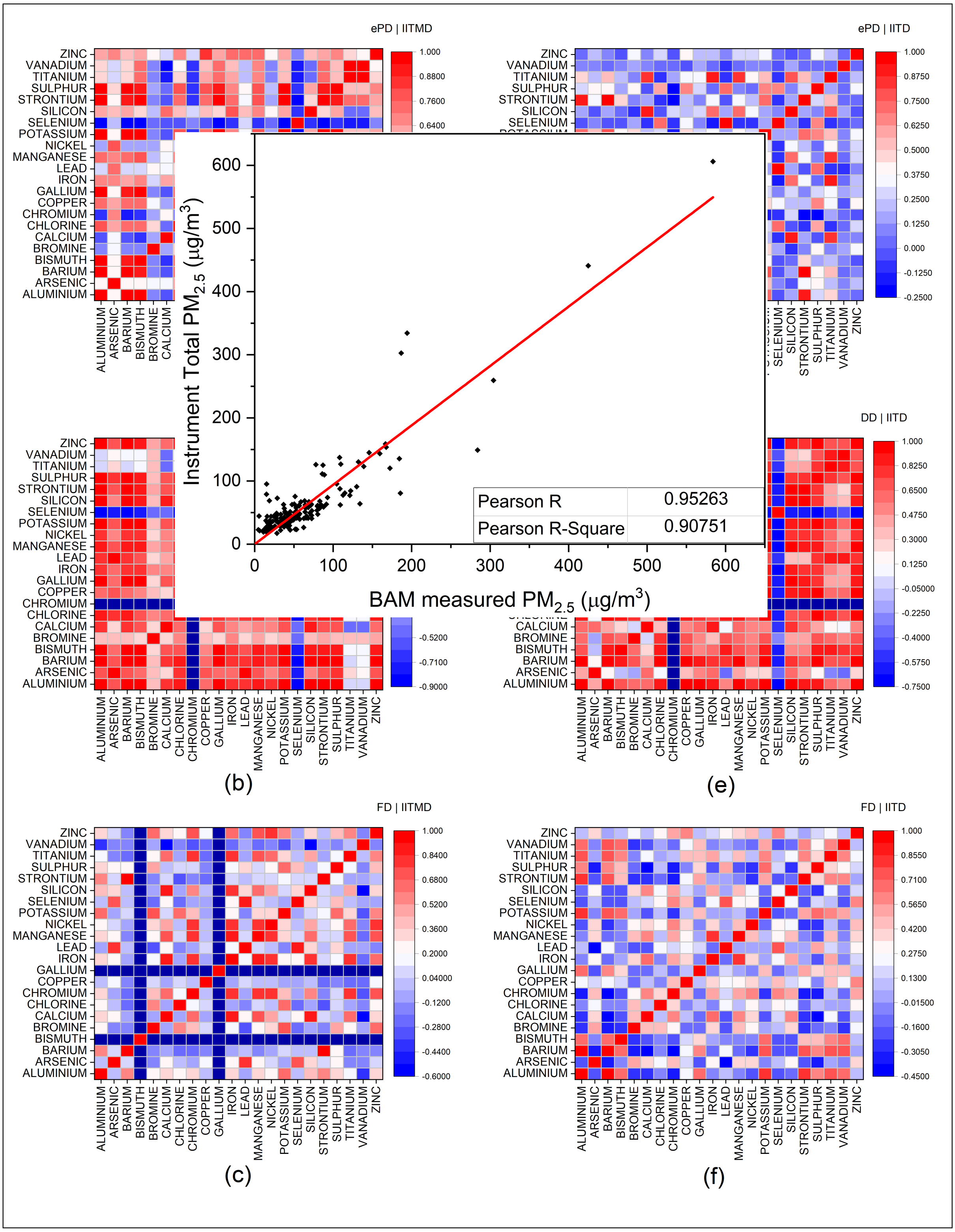

Figure S2: Pearson R Correlation statistics among the constituents of $PM_{2.5}^{el}$ measured at (a-c) IITMD through the phases (top to bottom) ePD, DD and FD; (d-f) IITD through the phases (top to bottom) ePD, DD, and FD

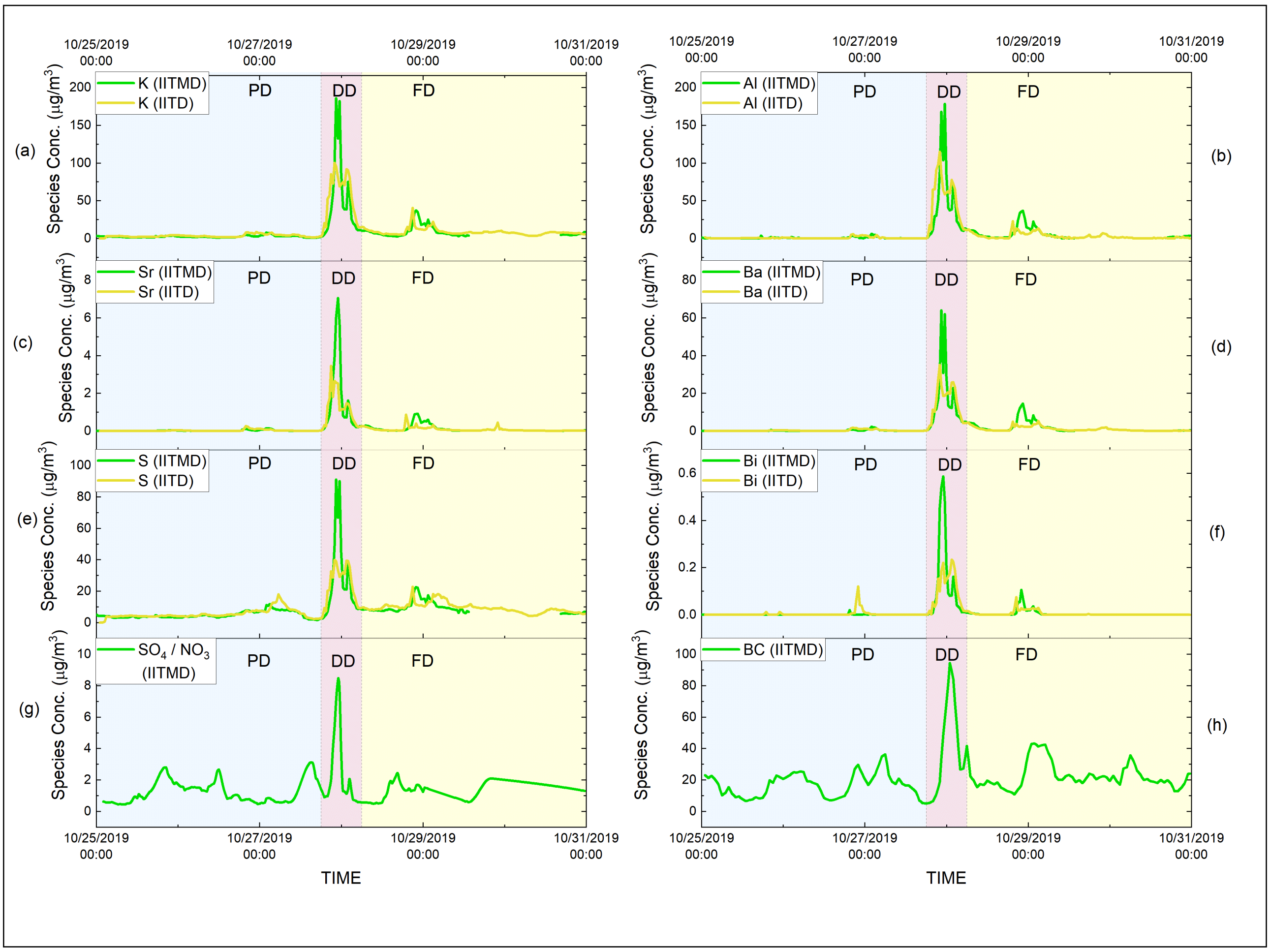

Figure S3: Temporal Variation of marker species for the fireworks event during Diwali at the IITMD (green) and IITD (yellow) sites; (a) Potassium; (b) Aluminum (c) Strontium; (d) Barium; (e) Sulfur; (f) Bismuth at the two sites using Xact 625i; (g) Sulfate to Nitrate ($SO_4/NO_3$) ratio measured using HR-ToF-AMS at IITMD site; (h) Black Carbon measured using Aethalometer AE33 at IITMD site

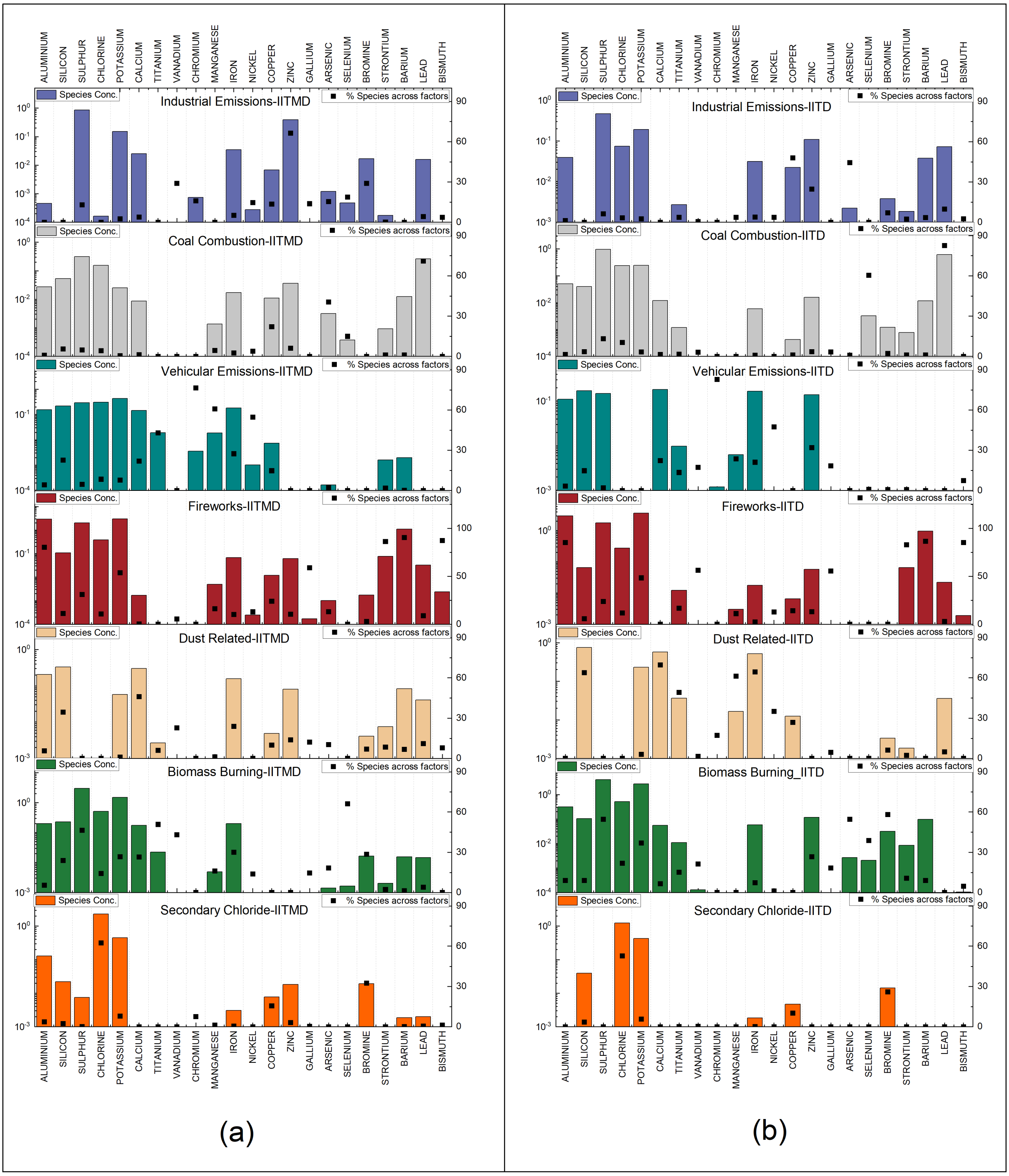

Figure S4: PMF resolved factor profiles; (top to bottom) industrial emissions, coal combustion, vehicular emissions, fireworks, dust-related, biomass burning, secondary chloride at (a) IITMD site; (b) IITD site

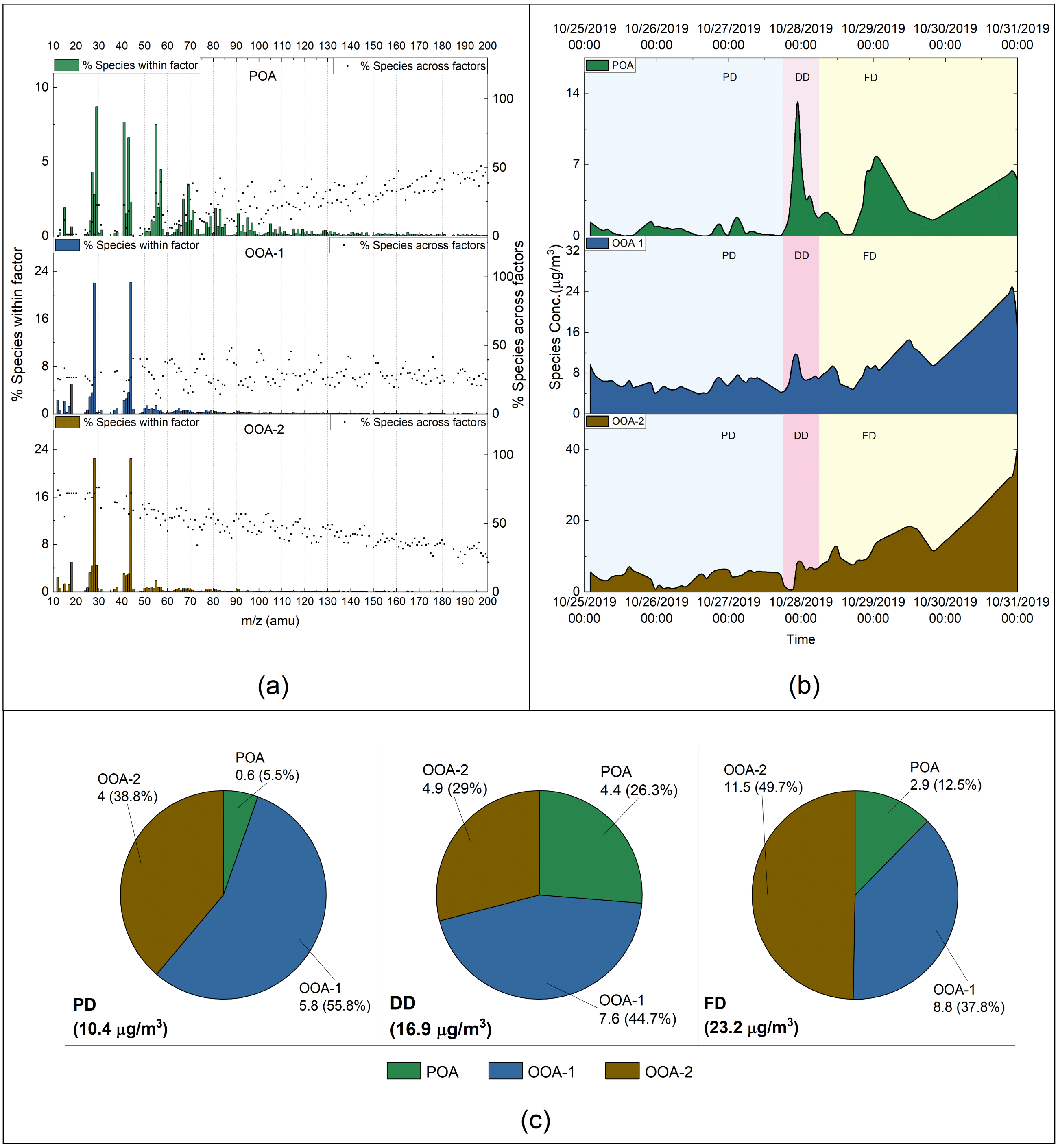

Figure S5: Source Apportionment Results of Organic $PM_{2.5}$ at IITMD (a) resolved factor profiles; (top to bottom) primary organic aerosol (POA), oxygenated organic aerosol - 1 (OOA-1), oxygenated organic aerosol – 2 (OOA-2); (b) temporal variation of each resolved factor; (top to bottom) primary organic aerosol (POA), oxygenated organic aerosol - 1 (OOA-1), oxygenated organic aerosol – 2 (OOA-2); (c) phase-wise composition of organic $PM_{2.5}$; (left to right) Pre-Diwali (PD), During Diwali (DD), Following Diwali (FD)

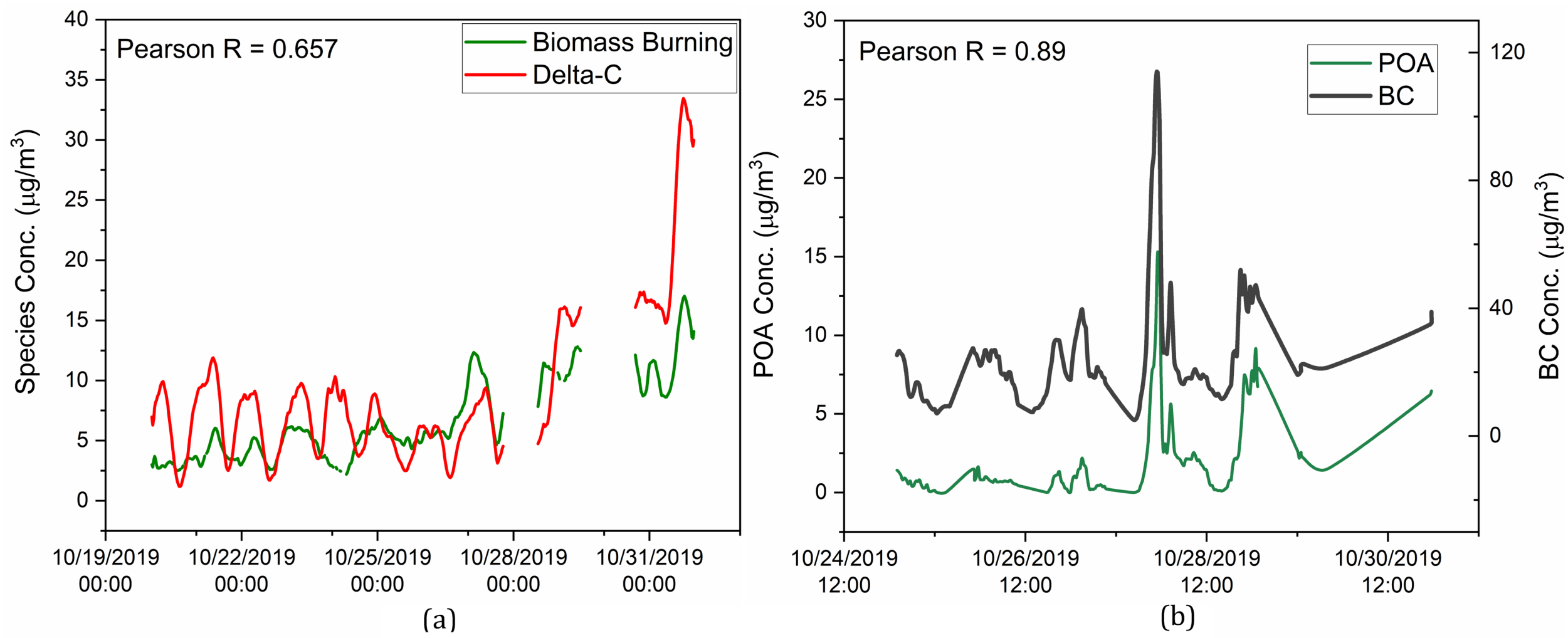

Figure S6: Correlation of PMF resolved sources with independent external markers (a) Xact-derived Biomass burning with Delta-C (from Aethalometer); (b) AMS derived primary organic aerosol (POA) with black carbon (from Aethalometer)

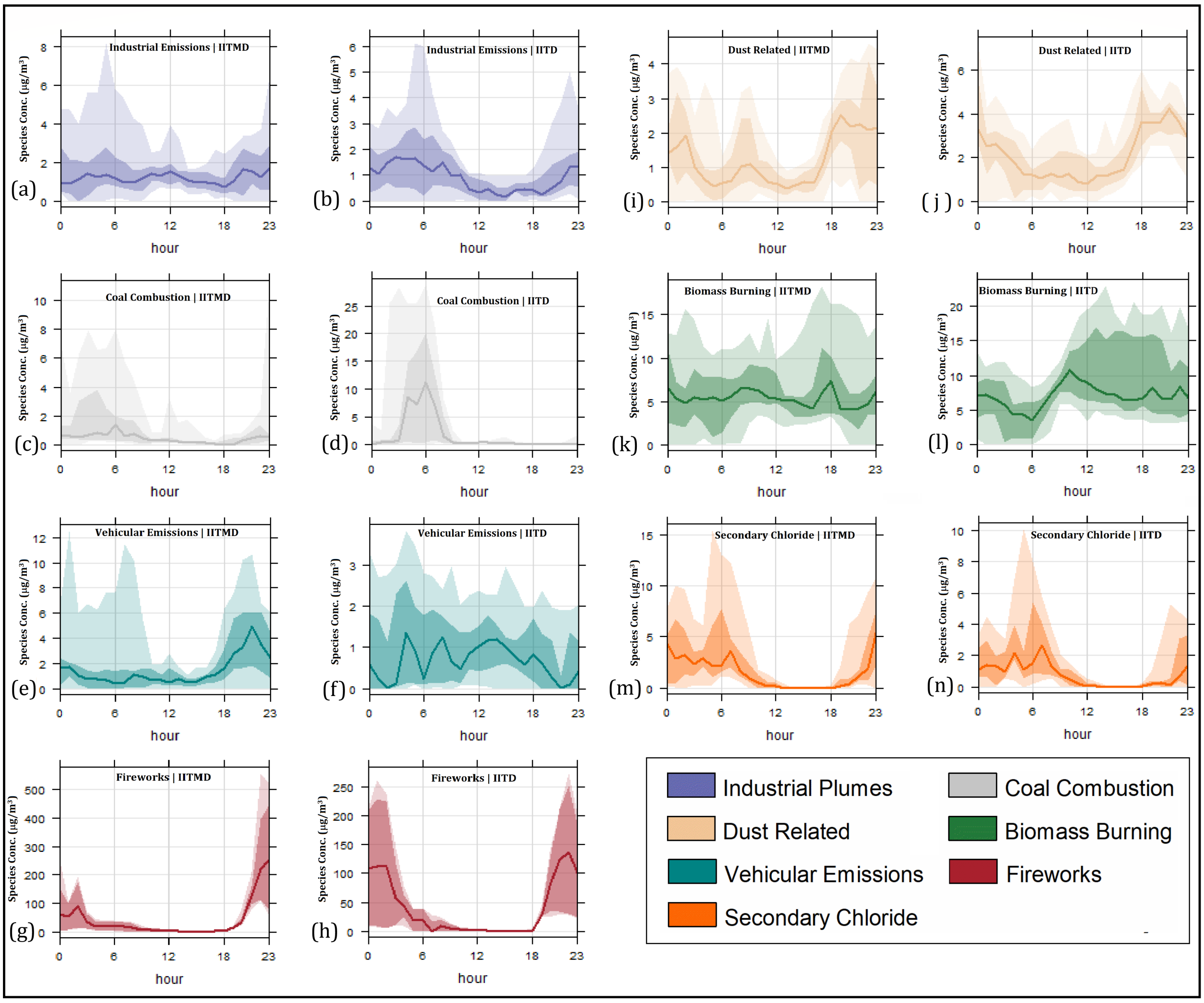

Figure S7: Diurnal Variation for sources resolved from apportionment of $PM_{2.5}^{el}$: (a,b) industrial emissions at (left to right) IITMD, IITD; (c,d) coal combustion at (left to right) IITMD, IITD; (e,f) vehicular emissions at (left to right) IITMD, IITD; (g,h) fireworks at (left to right) IITMD, IITD; (i,j) dust-related at (left to right) IITMD, IITD; (k,l) biomass burning at (left to right) IITMD, IITD; (m,n) secondary chloride at (left to right) IITMD, IITD

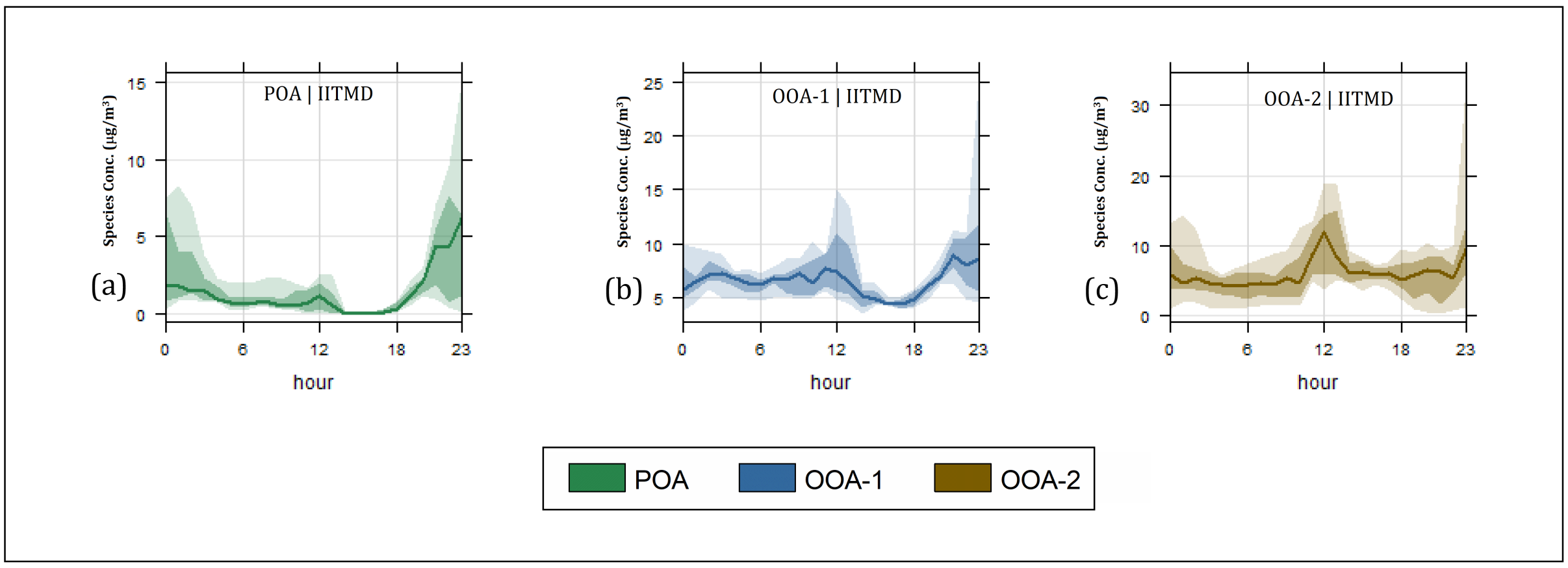

Figure S8: Diurnal Variation for sources resolved from apportionment of organic $PM_{2.5}$ measured at IITMD: (a) primary organic aerosols (POA); (b) oxygenated organic aerosol-1 (OOA-1); (c) oxygenated organic aerosol-2 (OOA-2)